\documentclass[12pt]{a_t}
\usepackage{graphicx}
\usepackage{cancel}
\usepackage{cite}
 \newtheorem{assumption}{\hspace{\parindent}\sl{A\,s\,s\,u\,m\,p\,t\,i\,o\,n\,}}

\begin{document}  
\numberwithin{equation}{section}
\def\figurename{Fig.}

\year{2022}
\title{Relaxation of Conditions for Convergence of Dynamic Regressor Extension and Mixing Procedure}%
\thanks{This research was in part financially supported by Grants Council of the President of the Russian Federation (project MD-1787.2022.4).}

\authors{A.I.~Glushchenko, Dr.Sc. (aiglush@ipu.ru),\\
K.A.~Lastochkin (lastconst@yandex.ru)\\
(Trapeznikov Institute of Control Sciences, Russian Academy of Sciences, Moscow, 117997 Russia)}

\maketitle

\begin{abstract}
A generalization of the dynamic regressor extension and mixing procedure is proposed, which, unlike the original procedure, first, guarantees a reduction of the unknown parameter identification error if the requirement of regressor semi-finite excitation is met, and second, it ensures exponential convergence of the regression function (regressand) tracking error to zero when the regressor is semi-persistently exciting with a rank one or higher.
\end{abstract}

\section{Introduction}
In recent years, in the literature on adaptive control and identification theory, more than a hundred papers have been published (see the references, the ones therein and review~\cite{1}) devoted to development of methods to identify unknown time-invariant parameters of linear regression equations with improved properties both in terms of transient quality indexes and the necessary conditions for parameters estimates convergence to their true values. A considerable part of these studies is based on the Dynamic Regressor Extension and Mixing (DREM) procedure~\cite{2} and its analogs (integral modification I-DREM~\cite{3}, procedures to generate new scalar excited regressor G+D and D+G~\cite{4, 5}, scalar identification schemes with finite-time convergence [6], etc.).

The basic DREM procedure~\cite{2} consists of the regressor extension and mixing steps. In the first step, the initial linear regression, which regressor is usually a vector, is transformed into an extended one with a square regressor matrix using stable dynamic operators and special extension schemes~\cite{1, 7, 8}. In the second step, the obtained equation is multiplied with the extended regressor adjunct matrix to convert it into a set of scalar equations with the same scalar regressor.

In contrast to the well-known conventional gradient identifier~\cite{9}, the DREM procedure~\cite{2}: 1) allows one to introduce a set of scalar estimation laws, each of which is responsible for identification of a certain unknown parameter, and the accuracy and convergence rate of such identification can be improved by adjustment of such laws scalar adaptive gains, and 2) relaxes the regressor persistent excitation requirement and guarantees asymptotic convergence of estimates to the true values if the scalar regressor is non-square integrable. Modified DREM procedures~\cite{3, 4, 5, 6}, in turn, relax this condition and ensure exponential or finite-time convergence of the parameter error to zero if the regressor is finitely or initially exciting.

However, as it has been analytically proved and experimentally demonstrated in~\cite{7, 8}, for DREM like procedures~\cite{2, 3, 4, 5, 6} the condition of the regressor finite excitation is necessary to obtain a scalar regressor that is bounded away from zero, and therefore it is a convergence condition. If this requirement is not met in schemes~\cite{2, 3, 4, 5, 6}, the unknown parameters identification error, as well as the regressand tracking error, cannot be reduced. At the same time, even when the condition of regression finite excitation is not satisfied, the classical gradient identifier~\cite{9} ensures the unknown parameters identification error reduction and an asymptotic convergence of the tracking error, which significantly narrows the applicability domain of the DREM-like procedures~\cite{2, 3, 4, 5, 6} in comparison with this approach.

Generally speaking, the condition of the regressor finite excitation is quite a weak requirement~\cite{10} and not satisfied in two main situations: 1) at least one element of the regressor is identically zero; 2) a linear dependence between the components of the regressor occurs~\cite{5}.

It is proved in~\cite{10} that the state vector of a stationary plant in the Frobenius form is excited finitely over the initial time interval if the reference signal is non-differentiable at least at one point of such interval, which is true, for example, if the reference signal is a Heaviside function. However, practical experience makes it possible to conclude that for each specific identification problem and each specific parameterization there exist their own particular requirements, which are necessary to ensure the regressor finite excitation. Currently, no generalized formalized criteria accepted by the control community have been proposed to verify {\it a priori} that the regressor is finitely exciting for an arbitrary parametrization. Therefore, as far as the identification and adaptive control problems are concerned, it is necessary to apply only such identification procedures and algorithms that are capable of ensuring the reduction of the unknown parameter estimation error and the convergence of the tracking error even when the regressor finite excitation is not provided, which, in particular, motivates the development of a modified dynamic regressor extension and mixing procedure with a relaxed convergence condition.

Such a relaxed requirement could be, for example, a semi-finite excitation condition, which, in contrast to the finite excitation condition, is met as long as at least one of the regressor elements is non-zero, even in case of linear dependence between all the regressor components~\cite{11}.

To date, two main approaches~\cite{12, 13, 14} have been proposed in the literature known to the authors that relax the convergence condition of the basic DREM procedure to the requirement of semi-finite excitation.

In~\cite{12}, an identification law with switches has been proposed, in which the I-DREM-based law is used when the condition of finite excitation is satisfied, and the conventional gradient law is applied when the requirement of semi-finite excitation is met. The main disadvantage of this approach is that in the second case it ensures the unknown parameter identification quality that coincides with the conventional gradient identifier. In~\cite{13, 14}, on the basis of the modified Gramm-Schmidt process, the algorithm to remove linearly dependent rows and columns from the extended regressor matrix has been developed, which allows one to reduce the problem of the unknown parameters identification to the problem of numerical solution of algebraic equations system in case the analytical dependence of the unknown parameters from each other is known. However, there are some hesitations that these equations can be solved when the unknown parameters are independent from each other and, consequently, the extension of the method from~\cite{13, 14} to the general case faces difficulties.

Thus, the problem to relax the convergence condition of the basic procedure of dynamic regressor extension and mixing is actual and does not have effective solutions up to date. Therefore, in this study a new step of regularization of the extended regressor is proposed to be added to the conventional DREM procedure to relax its convergence condition.

The aim of the regularization step is to, first, check the conditions that are necessary and sufficient to generate a scalar separated-from-zero regressor and, second, virtually change the matrix of the extended regressor when such conditions are violated. More specifically, in the regularization step we propose to apply the eigenvalue decomposition to the extended regressor obtained by the Kreisselmeyer filter~\cite{1}, which, because of such regressor symmetry and positive semi-definiteness, allows one to:
\begin{enumerate}
\item[--] verify that the condition of finite excitation of the extended regressor is met by analysis of its eigenvalues;
\item[--] following the ridge regression method~\cite{15, 16}, substitute  zero eigenvalues of the regressor with arbitrary constants.
\end{enumerate}

When the semi-finite excitation condition is met, mixing of the extended regressor modified by the regularization allows one to obtain a new regression with a non-zero scalar regressor over the semi-finite excitation time interval. Such result is impossible without regularization. In this study it is shown that the identification law based on such regression coincides with the DREM-based one if the regressor finite excitation requirement is met and, in addition, if the necessary condition of semi-finite excitation and a number of sufficient conditions are satisfied, it ensures both the identification and tracking errors decrease.

The main result of this research is a dynamic regressor extension, regularization, and mixing procedure that relaxes the convergence condition of the basic DREM method.

\begin{center}\it{Notation}\end{center}

The definitions from \cite{3, 9, 10, 11, 17}, which are used in axiomatic manner to state the problem and present the main result, are introduced.

\begin{definition} 
The regressor $\overline \varphi \left( t \right) \in {R^n}$ is persistently exciting $\left( {\overline \varphi \left( t \right) \in {\rm{PE}}} \right)$, \linebreak if $\forall t \ge {t_0} \ge 0$ $\exists T > 0$ and $\alpha  > 0$ such that the following holds
\begin{gather} \label{1.1}
{\lambda _{{\rm{min}}}}\left\{ {\int\limits_t^{t + T} {\overline \varphi \left( \tau  \right){{\overline \varphi }^{\rm{T}}}\left( \tau  \right)d} \tau } \right\} \ge \alpha ,
\end{gather}
where $\alpha  > 0$ is the excitation level, ${\lambda _{{\rm{min}}}}\left\{ . \right\}$ stands for the operator that returns the minimum eigenvalue of a matrix.
\end{definition}

\begin{definition} 
The regressor $\overline \varphi \left( t \right) \in {R^n}$ is finitely exciting $\left( {\overline \varphi \left( t \right) \in {\rm{FE}}} \right)$ over the time range $\left[ {t_r^ + {\rm{;\;}}{t_e}} \right] \subset \left[ {{t_0}{\rm{;\;}}\infty } \right)$, if there exist ${t_e} > t_r^ +  \ge {t_0} \ge 0$ and $\alpha  > 0$ such that
\begin{gather} \label{1.2}
{\lambda _{{\rm{min}}}}\left\{ {\int\limits_{t_r^ + }^{{t_e}} {\overline \varphi \left( \tau  \right){{\overline \varphi }^{\rm{T}}}\left( \tau  \right)d} \tau } \right\} \ge \alpha.
\end{gather}
\end{definition}

\begin{definition} 
The regressor $\overline \varphi \left( t \right) \in {R^n}$ is semi-persistently exciting $\left( {\overline \varphi \left( t \right) \in {\rm{s\text{-}PE}}} \right)$ with the time-invariant rank $0 < r < n$, if $\forall t > {t_0} \ge 0$ $\exists T > 0$ and $0 < \underline \alpha  \le \overline \alpha$ such that $\forall i \in \left\{ {1, \ldots {\rm{,}}r} \right\}$ the inequality holds
\begin{gather} \label{1.3}
 \underline \alpha \le {\lambda _i}\left\{ {\int\limits_t^{t + T} {\overline \varphi \left( \tau  \right){{\overline \varphi }^{\rm{T}}}\left( \tau  \right)d} \tau } \right\} \le \overline \alpha,
\end{gather}
where  $0 < \underline \alpha \le \overline \alpha $ is a partial excitation level.
\end{definition}

\begin{definition} 
The regressor $\overline \varphi \left( t \right) \in {R^n}$ is semi-finitely eciting $\left( {\overline \varphi \left( t \right) \in {\rm{s\text{-}FE}}} \right)$ with time-invariant rank $0 < r < n$ over the time range $\left[ {t_r^ + {\rm{;\;}}{t_e}} \right] \subset \left[ {{t_0}{\rm{;\;}}\infty } \right)$, if there exists ${t_e} > t_r^ +  \ge 0$ and $0 < \underline \alpha \le \overline \alpha$ such that $\forall i \in \left\{ {1, \ldots {\rm{,}}r} \right\}$ it holds that
\begin{gather} \label{1.4}
\underline \alpha \le {\lambda _i}\left\{ {\int\limits_{t_r^ + }^{{t_e}} {\overline \varphi \left( \tau  \right){{\overline \varphi }^{\rm{T}}}\left( \tau  \right)d} \tau } \right\} \le \overline \alpha.
\end{gather}
\end{definition}

The relations between the introduced regressor excitation types are specified as follows:
\begin{gather*}
\overline \varphi \left( t \right) \in {\rm{PE}} \Rightarrow \left\{ \begin{array}{c}
\overline \varphi \left( t \right) \in {\rm{FE}}\\
\overline \varphi \left( t \right) \in {\rm{s\text{-}PE}}
\end{array} \right\} \Rightarrow \overline \varphi \left( t \right) \in {\rm{s\text{-}FE}}{\rm{.}}
\end{gather*}

The requirements~\eqref{1.1} and~\eqref{1.2} impose constraints on all eigenvalues of the Gramm matrix, whereas~\eqref{1.3} and~\eqref{1.4} restrict only some of them. That is why the condition $\overline \varphi \left( t \right) \in {\rm{s\text{-}FE}}$ is the weakest one and, as far as limiting case is considered, is met when $r = 1$ if at least one of the regressor $\overline \varphi \left( t \right)$ elements is non-zero over the time range $\left[ {t_r^ + {\rm{;\;}}{t_e}} \right] \subset \left[ {{t_0}{\rm{;\;}}\infty } \right)$.

An important role in the modern identification theory is played by the Kreisselmeyer filtering, which allows one to transform a vector regressor $\overline \varphi \left( t \right) \in {R^n}$ into a matrix one $\varphi \left( t \right) \in {R^{n \times n}}$ without loss of excitation:
\begin{gather} \label{1.5}
\forall t \ge {t_0}{\rm{\;}}\dot \varphi \left( t \right) =  - l\varphi \left( t \right) + \overline \varphi \left( t \right){\overline \varphi ^{\rm{T}}}\left( t \right){\rm{,\;}}\varphi \left( {{t_0}} \right) = {0_{n \times n}}{\rm{,}}
\end{gather}
where $l > 0$ is the Kreisselmeyer filter parameter.

The properties of the matrix regressor $\varphi \left( t \right) \in {R^{n \times n}}$ with respect to the conditions (\ref{1.1})--(\ref{1.4}) on $\overline \varphi \left( t \right) \in {R^n}$ are copied from \cite{8, 11}.

\begin{corollary} 
$\overline \varphi \left( t \right) \in {\rm{PE}} \Leftrightarrow \forall t \ge kT{\rm{\;}}{\lambda _{{\rm{min}}}}\left( t \right) > \mu {\rm{.}}$
\end{corollary}

\begin{corollary} 
$\overline \varphi \left( t \right) \in {\rm{FE}} \Leftrightarrow \forall t \in \left[ {{t_\delta }{\rm{;\;}}{t_\delta } + \delta } \right] \subset \left[ {t_r^ + {\rm{;\;}}{t_e}} \right]{\rm{\;}}{\lambda _{{\rm{min}}}}\left( t \right) > \mu .$
\end{corollary}

\begin{corollary} 
$\overline \varphi \left( t \right) \in {\rm{s\text{-}PE}} \Leftrightarrow \forall t \ge kT{\rm{\;}}\forall i \in \left\{ {1, \ldots {\rm{,}}r} \right\}{\rm{\;}}{\lambda _i}\left( t \right) > \mu .$
\end{corollary}

\begin{corollary} 
$\overline \varphi \left( t \right) \in {\rm{s\text{-}FE}} \Leftrightarrow \forall t \in \left[ {{t_\delta }{\rm{;\;}}{t_\delta } + \delta } \right]\subset\left[ {t_r^ + {\rm{;}}{t_e}} \right]{\rm{}}\forall i \in \left\{ {1, \ldots {\rm{,}}r} \right\}{\rm{}}{\lambda _i}\left( t \right) > \mu .$
\end{corollary}

Here $k \ge 1$ is a positive integer number, $\mu  > 0$ is a lower bound of the eigenvalue, ${\lambda _i}\left( t \right)$ is the $i^{\rm th}$ eigenvalue of the regressor $\varphi \left( t \right)$,  ${\lambda _{{\rm{min}}}}\left( t \right) = \mathop {{\rm{min}}}\limits_{1 \le i \le n - \overline r} {\lambda _i}\left( t \right)$ is the minimum separated-from-zero eigenvalue of the regressor $\varphi \left( t \right)$, $\overline r = n - r$ is the rank deficiency.

The proofs of Corollary 1 and 2 are given in \cite{8, 11} respectively, while the proofs of Corollary 3 and 4 can be obtained in the same way.

Based on the definition of the eigenvalue decomposition of the positive semi-definite time-invariant matrix from \cite{17}, the definition of the eigenvalue decomposition of the dynamic regressor $\varphi \left( t \right) \in {R^{n \times n}}$ is introduced.

\begin{definition} 
The eigenvalue decomposition of the regressor $\varphi \left( t \right) \in {R^{n \times n}}$ with piecewise-constant rank $r\left( t \right) \le n$ is defined as follows:
\begin{gather} \label{1.6}
\begin{array}{c}
{V^{\rm{T}}}\left( t \right)\varphi \left( t \right)V\left( t \right) = \left[ {\begin{array}{*{20}{c}}
{V_1^{\rm{T}}\left( t \right)}\\
{V_2^{\rm{T}}\left( t \right)}
\end{array}} \right]\varphi \left( t \right)\left[ {\begin{array}{*{20}{c}}
{{V_1}\left( t \right)}&{{V_2}\left( t \right)}
\end{array}} \right] = \\
= \Lambda \left( t \right) = \left[ {\begin{array}{*{20}{c}}
{{\Lambda _1}\left( t \right)}&{{0_{r\left( t \right) \times \overline r\left( t \right)}}}\\
{{0_{\overline r\left( t \right) \times r\left( t \right)}}}&{{0_{\overline r\left( t \right)}}}
\end{array}} \right]{\rm{,}}\\
{\Lambda _1}\left( t \right) \in {R^{r\left( t \right) \times r\left( t \right)}} = {\rm{diag}}\left\{ {{\lambda _1}\left( t \right),{\lambda _2}\left( t \right), \ldots ,{\lambda _{r\left( t \right)}}\left( t \right)} \right\}{\rm{,}}
\end{array}
\end{gather}
where ${V_1}\left( t \right) \in {R^{n \times r\left( t \right)}}$ stands for a time-varying orthonormal basis of $\varphi \left( t \right)$ eigenspace, ${V_2}\left( t \right) \in {R^{n \times \overline r\left( t \right)}}$ is a time-varying orthonormal basis of $\varphi \left( t \right)$ nullspace, ${\lambda _1}\left( t \right) \ge {\lambda _2}\left( t \right) \ge  \cdots  \ge {\lambda _{r\left( t \right)}}\left( t \right) > 0$ denote nonzero eigenvalues of $\varphi \left( t \right)$, ${0_{\overline r\left( t \right)}} \in {R^{\overline r\left( t \right) \times \overline r\left( t \right)}}$ is a zero matrix, ${0_{\overline r\left( t \right) \times r\left( t \right)}} \in {R^{\overline r\left( t \right) \times r\left( t \right)}},{\rm{\;}}{0_{r\left( t \right) \times \overline r\left( t \right)}} \in {R^{r\left( t \right) \times \overline r\left( t \right)}}$ stands for zero matrices of corresponding dimensions.
\end{definition}

\section{Problem statement}

The classical problem of the time-invariant parameters identification of a linear regression equation is considered:
\begin{gather} \label{2.1}
\forall t \ge {t_0}{\rm{\;}}z\left( t \right) = {\overline \varphi ^{\rm{T}}}\left( t \right)\theta {\rm{,}}
\end{gather}
where $\overline \varphi \left( t \right) \in {R^n}$,  $z\left( t \right) \in R$ are measurable regressor and function (regressand), $\theta  \in {R^n}$ is a vector of unknown time-invariant $\left( {\dot \theta  \equiv 0} \right)$ and bounded $\left( {\left\| \theta  \right\| \le {\theta _{{\rm{max}}}}} \right)$ parameters.

It is assumed that the following assumption holds for $\overline \varphi \left( t \right)$.

\begin{assumption}
The regressor $\overline \varphi \left( t \right)$ is bounded: $\left\| {\overline \varphi \left( t \right)} \right\| \le {\overline \varphi _{{\rm{max}}}}$.
\end{assumption}

In general case, the above-stated requirement can be met with the help of multiplication of~\eqref{2.1} by ${n_s} = {\textstyle{1 \over {1 + {{\overline \varphi }^{\rm{T}}}\left( t \right)\overline \varphi \left( t \right)}}}$.

The aim is to derive the adaptive law to obtain the estimations $\hat \theta \left( t \right) \in {R^n}$, which, when $\overline \varphi \left( t \right) \in {\rm{s \text{-} FE}}$, ensures that:
\begin{gather} \label{2.2}
\begin{array}{c}
\left\| {\tilde \theta \left( {{t_e}} \right)} \right\| \le \beta \left\| {\tilde \theta \left( {t_r^ + } \right)} \right\|{\rm{,\;}}\beta  \in \left( {0{\rm{; 1}}} \right){\rm{,}}\\
\left| {\tilde z\left( {{t_e}} \right)} \right| \le \beta \left| {\tilde z\left( {t_r^ + } \right)} \right|{\rm{,}}
\end{array}
\end{gather}
where $\tilde z\left( t \right) = {\overline \varphi ^{\rm{T}}}\left( t \right)\hat \theta  - z\left( t \right)$ is the tracking error, $\tilde \theta \left( t \right) = \hat \theta \left( t \right) - \theta $ stands for the parameter error.

The inequalities~\eqref{2.2} mean the reduction of $\tilde \theta \left( t \right)$ and $\tilde z\left( t \right)$ respectively over the time range $\left[ {t_r^ + {\rm{;\;}}{t_e}} \right]$. The requirement $\overline \varphi \left( t \right) \in {\rm{s \text{-} FE}}$ is the convergence condition of the desired adaptive law. The convergence is capability to reduce the initial values of the errors $\tilde z\left( {t_r^ + } \right)$ and $\tilde \theta \left( {t_r^ + } \right)$.

\subsection{Gradient-based identification law}

The classical solution, which ensures that the goal~\eqref{2.2} is achieved, is the gradient-based identification law:
\begin{gather} \label{2.3}
\dot {\hat {\theta}} \left( t \right) =  - \Gamma \overline \varphi \left( t \right){\left( {\overline \varphi^{\rm{T}}\left( t \right){{\hat \theta }}\left( t \right) - z\left( t \right)} \right)}{\rm{,\;}}\Gamma  = {\Gamma ^{\rm{T}}} > 0,
\end{gather}
which convergence is guaranteed when $\overline \varphi \left( t \right) \in {\rm{s \text{-} FE}}$ and, in general, it ensures the following properties:
\begin{enumerate}
  \setlength{\leftskip}{15pt}
\item[$a_1)$] $\overline \varphi \left( t \right) \in {\rm{PE}} \Leftrightarrow \left\{ \begin{array}{l}
{\rm{li}}{{\rm{m}}_{t \to \infty }}\left\| {\tilde \theta \left( t \right)} \right\| = 0{\rm{ }}\left( {{\rm{exp}}} \right)\\
{\rm{li}}{{\rm{m}}_{t \to \infty }}\left| {\tilde z\left( t \right)} \right| = 0{\rm{ }}\left( {{\rm{exp}}} \right)
\end{array} \right.{\rm{;}}$
\item[$a_2)$] ${\rm{li}}{{\rm{m}}_{t \to \infty }}\left| {\tilde z\left( t \right)} \right| = 0{\rm{;}}$
\item[$a_3)$] ${\lambda _{{\rm{min}}}}\left( \Gamma  \right) = {\lambda _{{\rm{max}}}}\left( \Gamma  \right) \Rightarrow \left\| {\tilde \theta \left( {{t_a}} \right)} \right\| \le \left\| {\tilde \theta \left( {{t_b}} \right)} \right\|{\rm{ }}\forall {t_a} \ge {t_b}{\rm{;}}$
\item[$a_4)$]$\overline \varphi \left( t \right) \in {\rm{s \text{-} FE}} \Rightarrow \left\{ \begin{array}{l}
\left\| {\tilde \theta \left( {{t_e}} \right)} \right\| \le \beta \left\| {\tilde \theta \left( {t_r^ + } \right)} \right\|{\rm{,\;}}\beta  \in \left( {0{\rm{;\;1}}} \right)\\
\left| {\tilde z\left( {{t_e}} \right)} \right| \le \beta \left| {\tilde z\left( {t_r^ + } \right)} \right|\end{array} \right.{\rm{;}}$
\item[$a_5)$] when $\overline \varphi \left( t \right) \in {\rm{PE}}$ there is an optimal value of $\Gamma$ that maximizes the rate of exponential convergence of the parameter error $\tilde \theta \left( t \right)$ to zero. The change of any element of the matrix $\Gamma$ affects the transients quality of all ${\tilde \theta _i}\left( t \right)$.
\end{enumerate}

Despite ensuring some properties when $\overline \varphi \left( t \right) \in {\rm{s\text{-} FE}}$ ($a_4$), the law \eqref{2.3} guarantees exponential convergence of $\tilde \theta \left( t \right)$ and $\tilde z\left( t \right)$ to zero if the strict condition of the regressor persistent excitation ($a_1$) is met, provides monotonicity of the parameter error norm only ($a_3$), and each element of the arbitrary parameter $\Gamma$ affects the transients quality of all errors ($a_5$).

To overcome the disadvantages of the law~\eqref{2.3} in~\cite{2}, a DREM procedure has been proposed, according to which, firstly, the regression (\ref{2.1}) is processed using the regressor extension and mixing operations, and then, on the basis of the obtained new regression, the unknown parameter identification law is introduced. The synthesis procedure and properties of such a law is considered below.

\subsection{Dynamic regressor extension and mixing}

In the step of extension the initial vector regressor $\overline \varphi \left( t \right) \in {R^n}$ is transformed into the matrix one $\varphi \left( t \right) \in {R^{n \times n}}$ using, for an instance, the filter \eqref{1.5}:
\begin{gather} \label{2.4}
\begin{array}{l}
\dot \varphi \left( t \right) =  - l\varphi \left( t \right) + \overline \varphi \left( t \right){{\overline \varphi }^{\rm{T}}}\left( t \right){\rm{,\;}}\varphi \left( {{t_0}} \right) = {0_{n \times n}}{\rm{,}}\\
\dot y\left( t \right) =  - ly\left( t \right) + \overline \varphi \left( t \right)z\left( t \right){\rm{,\;}}{\mathop{\ y}\nolimits} \left( {{t_0}} \right) = {0_n}{\rm{,}}
\end{array}
\end{gather}
where $y\left( t \right) \in {R^n}$ is the extended regressand.

After filtering \eqref{2.4} the extended regression equation is obtained:
\begin{gather} \label{2.5}
y\left( t \right) = \varphi \left( t \right)\theta ,
\end{gather}
which regressor, in accordance with Corollary 1-4, could be used to verify the fact that any of the conditions \eqref{1.1}--\eqref{1.4} is met.

In the mixing step, in accordance with~\cite{2}, the matrix regressor $\varphi \left( t \right) \in {R^{n \times n}}$ is transformed into scalar one $\omega \left( t \right) \in R$ by way of multiplication of \eqref{2.5} by the adjoint matrix ${\rm{adj}}\left\{ {\varphi \left( t \right)} \right\}$ and application of the property ${\rm{adj}}\left\{ {\varphi \left( t \right)} \right\}\varphi \left( t \right) = {\rm{det}}\left\{ {\varphi \left( t \right)} \right\}{I_{n \times n}}$:
\begin{gather} \label{2.6}
\begin{array}{c}
Y\left( t \right) = \omega \left( t \right)\theta {\rm{,}}\\
Y\left( t \right){\rm{:}} = {\rm{adj}}\left\{ {\varphi \left( t \right)} \right\}y\left( t \right){\rm{,\;}}\omega \left( t \right){\rm{:}} = {\rm{det}}\left\{ {\varphi \left( t \right)} \right\}{\rm{,}}
\end{array}
\end{gather}
where $Y\left( t \right) \in {R^n}$.

On the basis of the obtained $n$ scalar equations \eqref{2.6} the following identification law is introduced according to~\cite{2}:
\begin{gather} \label{2.7}
{\dot {\hat {\theta _i}}}\left( t \right) = {\dot {\tilde{ \theta _i}}}\left( t \right) =  - {\gamma _i}\omega \left( t \right)\left( {\omega \left( t \right){{\hat \theta }_i}\left( t \right) - \omega \left( t \right){\theta _i}\left( t \right)} \right) =  - {\gamma _i}{\omega ^2}\left( t \right){\tilde \theta _i}\left( t \right){\rm{,\;}}{\gamma _i} > 0,
\end{gather}
which convergence condition is $\overline \varphi \left( t \right) \in {\rm{FE}}$, and it ensures the following properties:
\begin{enumerate}
  \setlength{\leftskip}{15pt}
\item[$b_1)$] $\begin{array}{l}
\omega \left( t \right) \notin {L_2} \Leftrightarrow {\rm{li}}{{\rm{m}}_{t \to \infty }}\left\| {\tilde \theta \left( t \right)} \right\| = 0{\rm{;}}\\
\omega \left( t \right) \in {\rm{PE}} \Leftrightarrow {\rm{li}}{{\rm{m}}_{t \to \infty }}\left\| {\tilde \theta \left( t \right)} \right\| = 0{\rm{ }}\left( {{\rm{exp}}} \right){\rm{;}}
\end{array}$
\item[$b_2)$] ${\rm{li}}{{\rm{m}}_{t \to \infty }}\left\| {\tilde \theta \left( t \right)} \right\| = 0 \Rightarrow {\rm{li}}{{\rm{m}}_{t \to \infty }}\underbrace {\left| {z\left( t \right) - {{\overline \varphi }^{\rm{T}}}\left( t \right)\hat \theta \left( t \right)} \right|}_{\left| {\tilde z\left( t \right)} \right|} = 0{\rm{\;}}\left( {certainty} {\rm{\;}} {equialence} \right){\rm{;}}$
\item[$b_3)$] $\left| {{{\tilde \theta }_i}\left( {{t_a}} \right)} \right| \le \left| {{{\tilde \theta }_i}\left( {{t_b}} \right)} \right|{\rm{ }}\forall {t_a} \ge {t_b}$;
\item[$b_4)$]$\overline \varphi \left( t \right) \in {\rm{FE}} \Rightarrow \left\{ \begin{array}{l}
\left\| {\tilde \theta \left( {{t_e}} \right)} \right\| \le \beta \left\| {\tilde \theta \left( {t_r^ + } \right)} \right\|{\rm{,\;}}\beta  \in \left( {0{\rm{; 1}}} \right)\\
\left| {\tilde z\left( {{t_e}} \right)} \right| \le \beta \left| {\tilde z\left( {t_r^ + } \right)} \right|
\end{array} \right.{\rm{;}}$
\item[$b_5)$] when $\overline \varphi \left( t \right) \in {\rm{PE}}$, the exponential convergence rate of the parameter error ${\tilde \theta _i}\left( t \right)$ can be improved by increase of $\gamma_i$, and change of any element $\gamma_i$ affects only the transient quality of the respective ${\tilde \theta _i}\left( t \right)$.
\end{enumerate}

As follows from the comparison of the properties $a_1$-$a_5$ and $b_1$-$b_5$, the relaxed requirement of asymptotic convergence of the parameter error ($b_1$), the monotonicity of the transients of each particular error ${\tilde \theta _i}\left( t \right)$ ($b_3$) as well as the fact that the transients quality of estimates for each particular ${\tilde \theta _i}\left( t \right)$ ($b_5$) can be adjusted with the help of ${\gamma _i}$ are the advantages of \eqref{2.7} compared to the gradient law \eqref{2.3}. However, at the same time, the law \eqref{2.7} does not provide convergence to zero of the error $\tilde z\left( t \right)$ separately from the parameter error convergence ($b_2$) and has a stricter convergence condition ($b_4$).

Therefore, the main goal of this study is to develop an identification law that combines the positive properties of \eqref{2.3} and \eqref{2.7}, which means that, when $\overline \varphi \left( t \right) \in {\rm{FE}}$, the proposed law is required to have the properties $b_1$-$b_5$ of \eqref{2.7}, while when $\overline \varphi \left( t \right) \in {\rm{s \text{-} FE}}$ -- the property $a_4$ of the law \eqref{2.3}, and in contrast to \eqref{2.7} it is required to ensure the convergence of the tracking error $\tilde z\left( t \right)$ separately from the parameter error convergence ($a_2$).

\section{Main result}

\subsection{Dynamic regularization of extended regressor}

Following Definition 5, the regression equation \eqref{2.5} is rewritten as
\begin{gather} \label{3.1}
y\left( t \right) = \varphi \left( t \right)\theta  = \left[ {\begin{array}{*{20}{c}}
{{V_1}\left( t \right)}&{{V_2}\left( t \right)}
\end{array}} \right]\Lambda \left( t \right)\left[ {\begin{array}{*{20}{c}}
{V_1^{\rm{T}}\left( t \right)}\\
{V_2^{\rm{T}}\left( t \right)}
\end{array}} \right]\theta  = V\left( t \right)\Lambda \left( t \right){V^{\rm{T}}}\left( t \right)\theta .
\end{gather}

It should be noted that, when ${\rm{rank}}\left\{ {\varphi \left( t \right)} \right\} = r\left( t \right) < n$, the matrix $\Lambda \left( t \right)$ contains $\overline r\left( t \right) > 0$ zeros on the main diagonal, and therefore $\omega\left( t \right)={\rm{det}}\left\{ {\varphi \left( t \right)} \right\} \equiv 0 \Rightarrow \left\| {\tilde \theta \left( {{t_e}} \right)} \right\| =\linebreak= \left\| {\tilde \theta \left( {t_r^ + } \right)} \right\|$. As a result, in order to make the regressor determinant $\varphi \left( t \right)$ be bounded away from zero when the regression with the scalar regressor \eqref{2.6} is obtained, the zeros of the main diagonal of the matrix $\Lambda \left( t \right)$ are to be virtually substituted with non-zero numbers~\cite{15, 16}. To achieve this, we introduce the matrix $\Xi \left( t \right)$, which being added to $\Lambda \left( t \right)$ allows one to obtain a full rank matrix:
\begin{gather} \label{3.2}
\begin{array}{c}
\Xi \left( t \right) = \overline \Lambda \left( t \right) - \Lambda \left( t \right){\rm{, }}\\
\overline \Lambda \left( t \right){\rm{:}} = \left\{ \begin{array}{l}
0_{n \times n},\text{\rm{ if diag}}\left\{ {{{\overline \lambda }_1}\left( t \right){\rm{,\;}}{{\overline \lambda }_2}\left( t \right){\rm{,}} \ldots {\rm{,}}{{\overline \lambda }_n}\left( t \right)} \right\} = \varepsilon {I_{n \times n}}\\
{\rm{diag}}\left\{ {{{\overline \lambda }_1}\left( t \right){\rm{,\;}}{{\overline \lambda }_2}\left( t \right){\rm{,}} \ldots {\rm{,}}{{\overline \lambda }_n}\left( t \right)} \right\}\text{\rm{, otherwise}}
\end{array} \right.{\rm{,\;}}i = \overline {1,n} {\rm{,\;}}\\
{{\overline \lambda }_i}\left( t \right){\rm{:}} = \left\{ \begin{array}{l}
{\lambda _i}\left( t \right)\text{\rm{, if }}{\lambda _i}\left( t \right) \ge \overline \varepsilon \\
\varepsilon \text{\rm{, if }}{\lambda _i}\left( t \right) < \overline \varepsilon 
\end{array} \right.{\rm{, }}
\end{array}
\end{gather}
where $\overline \Lambda \left( t \right)$ is a new matrix of eigenvalues, $\varepsilon  > 0$ stands for a parameter that defines the value of the virtual eigenvalues, $\overline \varepsilon  \ge 0$ denotes the parameter that defines the amplitude of the eigenvalues of  $\varphi \left( t \right)$, which are considered to be equivalently equal to zero in the presence of computation errors and external disturbances.

The expression $\pm V\left( t \right)\Xi \left( t \right){V^{\rm{T}}}\left( t \right)\theta$ is added to \eqref{3.1} to obtain:
\begin{gather} \label{3.3}
\begin{array}{c}
y\left( t \right) = \varphi \left( t \right)\theta  = V\left( t \right)\Lambda \left( t \right){V^{\rm{T}}}\left( t \right)\theta  \pm V\left( t \right)\Xi \left( t \right){V^{\rm{T}}}\left( t \right)\theta  = \\
 = V\left( t \right)\overline \Lambda \left( t \right){V^{\rm{T}}}\left( t \right)\theta  - V\left( t \right)\Xi \left( t \right){V^{\rm{T}}}\left( t \right)\theta  = \Phi \left( t \right)\theta  - V\left( t \right)\Xi \left( t \right){V^{\rm{T}}}\left( t \right)\theta, 
\end{array}
\end{gather}
where $\Phi \left( t \right) \in {R^{n \times n}}$ is a new regressor with the eigenvalues $\overline \Lambda \left( t \right)$.

The equation \eqref{3.3} is multiplied by the matrix ${\rm{adj}}\left\{ {\Phi \left( t \right)} \right\}$, and then the following properties are applied:
\begin{gather*}
\begin{array}{c}
{\rm{adj}}\left\{ {\Phi \left( t \right)} \right\} = {\rm{det}}\left\{ {\Phi \left( t \right)} \right\}{\Phi ^{ - 1}}\left( t \right){\rm{,\;}}{\Phi ^{ - 1}}\left( t \right) = V\left( t \right){{\overline \Lambda }^{ - 1}}\left( t \right){V^{\rm{T}}}\left( t \right){\rm{,\;}}\\
{\rm{adj}}\left\{ {\Phi \left( t \right)} \right\}\Phi \left( t \right) = {\rm{det}}\left\{ {\Phi \left( t \right)} \right\}{I_n}{\rm{,}}
\end{array}
\end{gather*}
to obtain:
\begin{gather} \label{3.4}
\begin{array}{c}
\Upsilon \left( t \right) = \omega \left( t \right)\theta  - \omega \left( t \right)V\left( t \right){{\overline \Lambda }^{ - 1}}\left( t \right)\Xi \left( t \right){V^{\rm{T}}}\left( t \right)\theta  = \omega \left( t \right)\Theta \left( t \right){\rm{,}}\\
\Upsilon \left( t \right){\rm{:}} = {\rm{adj}}\left\{ {\Phi \left( t \right)} \right\}y\left( t \right){\rm{,\;}}\omega \left( t \right){\rm{:}} = {\rm{det}}\left\{ {\Phi \left( t \right)} \right\}{\rm{, }}\\
\Theta \left( t \right){\rm{:}} = \theta  - V\left( t \right){{\overline \Lambda }^{ - 1}}\left( t \right)\Xi \left( t \right){V^{\rm{T}}}\left( t \right)\theta  = \theta  - \underbrace {{V_2}\left( t \right)V_2^{\rm{T}}\left( t \right)\theta }_{d\left( t \right)},
\end{array}
\end{gather}
where $\Theta \left( t \right) \in {R^n}$ is a vector of new unknown parameters, $d\left( t \right) \in {R^n}$ is a disturbance, which causes the difference between $\Theta \left( t \right)$ and $\theta$.

The properties of the new regressor $\omega \left( t \right) \in R$ are presented in the following proposition.
\begin{proposition}\label{st1}
Let the matrix $\overline \Lambda \left( t \right)$ be obtained using the equation \eqref{3.2} in case $\overline \varepsilon  = 0$, then the following implications hold:
\begin{enumerate}
  \setlength{\leftskip}{15pt}
\item[$1)$] $\overline \varphi \left( t \right) \in {\rm{PE}} \Leftrightarrow \forall t \ge kT{\rm{\;}}\omega \left( t \right) \ge \lambda _{\min }^n\left( t \right) > {\mu ^n} > 0.$
\item[$2)$]$\overline \varphi \left( t \right) \in {\rm{FE}} \Leftrightarrow \forall t \in \left[ {{t_\delta }{\rm{;\;}}{t_\delta } + \delta } \right] \subset \left[ {t_r^ + {\rm{;\;}}{t_e}} \right]{\rm{\;}}\omega \left( t \right) \ge \lambda _{\min }^n\left( t \right) > {\mu ^n} > 0.$
\item[$3)$]$\overline \varphi \left( t \right) \in {\rm{s \text{-} PE}} \Leftrightarrow \forall t \ge kT{\rm{\;}}\omega \left( t \right) \ge {\rm{min}}\left\{ {\lambda_{{\rm{min}}}^n\left( t \right){\rm{,\;}}{\varepsilon ^n}} \right\} > 0.$
\item[$4)$]$\overline \varphi \left( t \right) \in {\rm{s\text{-} FE}} \Leftrightarrow \forall t \in \left[ {{t_\delta }{\rm{;\;}}{t_\delta } + \delta } \right] \subset \left[ {t_r^ + {\rm{;\;}}{t_e}} \right]{\rm{\;}}\omega \left( t \right) \ge {\rm{min}}\left\{ {\lambda _{{\rm{min}}}^n\left( t \right){\rm{,\;}}{\varepsilon ^n}} \right\} > 0.$
\end{enumerate}
\end{proposition}

Proof of Proposition 1 is postponed to Appendix.

~

Using the regression \eqref{3.4} and the properties proved in Proposition 1, the identification law with normalization of the regressor excitation is introduced in accordance with~\cite{18}:
\begin{gather} \label{3.5}
\begin{array}{c}
\dot {\hat{ \theta}} \left( t \right) =  - \gamma \left( t \right)\omega \left( t \right)\left( {\omega \left( t \right)\hat \theta \left( t \right) - \Upsilon \left( t \right)} \right) = \\
 =  - \gamma \left( t \right){\omega ^2}\left( t \right)\left( {\hat \theta \left( t \right) - \theta } \right) - \gamma \left( t \right){\omega ^2}\left( t \right)d\left( t \right) = \\ = - \gamma \left( t \right){\omega ^2}\left( t \right)\underbrace {\left( {\hat \theta \left( t \right) - \Theta \left( t \right)} \right)}_{\tilde \Theta \left( t \right)}{\rm{,\;}}\hat \theta \left( {t_r^ + } \right) = {\theta _0}{\rm{, }}\\
\gamma \left( t \right){\rm{:}} = \left\{ \begin{array}{l}
{\gamma _1}{\rm{,\;}} \text{if} \; \omega \left( t \right) \le {\rm{min}}\left\{ {\lambda _{{\rm{min}}}^n\left( t \right){\rm{,\;}}{\varepsilon ^n}} \right\}\\
{\textstyle{{{\gamma _0}} \over {{\omega ^2}\left( t \right)}}}{\rm{\;}} \text{otherwise}
\end{array} \right.,
\end{array}
\end{gather}
where ${\gamma _0} > 0,{\rm{\;}}{\gamma _1} > 0$ are arbitrary parameters of the identification law, $\tilde \Theta \left( t \right) \in {R^n}$ is the error of the vector $\Theta \left( t \right)$ identification.

Owing to the algorithm to form the matrix $\Xi \left( t \right)$, the following theorem is valid for the law \eqref{3.5}.
\begin{theorem} \label{th:1}
Let Assumption 1 be met and $\overline \varepsilon  = 0$, then:
\begin{enumerate}
 \setlength{\leftskip}{15pt}
    \item[$1)$]if ${{\overline \varphi \left( t \right) \in {\rm{FE}}}/\overline \varphi \left( t \right) \in {\rm{PE}}}$, then \eqref{3.5} has the properties $b_1$--$b_5$;
    \item[$2)$] if $\overline \varphi \left( t \right) \in {\rm{s \text{-} FE}}$ and the following sufficient conditions are met 
    \begin{enumerate}
     \setlength{\leftskip}{15pt}
        \item[$2.1)$] $\left\| {\tilde \theta \left( {t_r^ + } \right)} \right\| = {\beta _1}{\theta _{{\rm{max}}}}{\rm{,\;}}{\beta _1} > 1,$
        \item[$2.2)$] the multiplication ${\gamma _0}\delta$ is such that  ${\textstyle{1 \over {{\beta _1}}}} + {e^{ - 0,5{\gamma _0}\delta }} \in \left( {0{\rm{; 1}}} \right)$,
    \end{enumerate}
    \item[] then the inequalities \eqref{2.2} holds, and the convergence conditions of \eqref{3.5} are satisfied;
    \item[$3)$] $\omega \left( t \right) \notin {L_2} \Rightarrow {\rm{li}}{{\rm{m}}_{t \to \infty }}\left\| {\tilde \theta \left( t \right)} \right\| \le {\theta _{{\rm{max}}}}{\rm{;}}$
    \item[$4)$]$\overline \varphi \left( t \right) \in {\rm{s \text{-} PE}} \Rightarrow {\rm{li}}{{\rm{m}}_{t \to \infty }}\left\| {\tilde \theta \left( t \right)} \right\| \le {\theta _{{\rm{max}}}}\left( {{\rm{exp}}} \right).$
\end{enumerate}

In this case the rate of exponential convergence can be directly adjusted by value of the parameter ${\gamma _0}$.
\end{theorem}

Proof of Theorem 1 is given in Appendix.

~

As follows from the results of Theorem 1, unfortunately, the law \eqref{3.5} does not capable of achievement the goal \eqref{2.2} if the values of $\tilde \theta \left( {t_r^ + } \right)$ are chosen arbitrarily, because in a set with a bound $\theta _{{{\rm{max}}}}$ the error norm $\tilde \theta \left( t \right)$ could become greater than $\left\| {\tilde \theta \left( {t_r^ + } \right)} \right\|$, which is a disadvantage of the law \eqref{3.5} compared to the conventional gradient \eqref{2.3}. Therefore, the necessary condition for convergence of \eqref{3.5} is a semi-finite excitation of the regressor $\overline \varphi \left( t \right) \in {\rm{s \text{-} FE}}$, while the sufficient condition is that premises 2.1) and 2.2) are met. Here it should also be noted that the choice $\hat \theta \left( {t_r^ + } \right) = {0_n}$ guarantees that the error $\tilde \theta \left( t \right)$ does not increase over the time range $\left[ t_r^+ {\rm{;\;}}{t_e} \right]$. So it can be stated that the law \eqref{3.5} is quasi-convergent in terms of \eqref{2.2} when only the necessary condition $\overline \varphi \left( t \right) \in {\rm{s \text{-} FE}}$ is satisfied.

Thus, according to the proposed dynamic regressor regularization procedure \eqref{3.1}-\eqref{3.4}, on the one hand, when $\overline \varphi \left( t \right) \in {\rm{FE}}$, the matrix $\Lambda \left( t \right)$ is not added with $\Xi \left( t \right)$ to form a full-rank matrix, and the law \eqref{3.5} reduces to \eqref{2.7}, and on the other hand, when $\overline \varphi \left( t \right) \in {\rm{s \text{-} FE}}$, the matrix $\Lambda \left( t \right)$ is added with $\Xi \left( t \right)$ to form a full rank matrix, and, in contrast to \eqref{2.7}, \eqref{3.5} ensures convergence in terms of \eqref{2.2} if the sufficient conditions are satisfied.

When the law \eqref{3.5} is applied, the global stability of the errors $\tilde z\left( t \right)$ and $\tilde \Theta \left( t \right)$ is analyzed by making different assumptions about the rank $r\left( t \right)$ and the basis of the nullspace ${V_2}\left( t \right)$. In subsection 3.2 it is assumed that they are time-invariant, whereas in section 3.3 they are considered to be piecewise-constant functions.

\subsection{Time-invariant rank and basis of nullspace}

The following assumption about the time-invariance of the rank and nullspace basis of the regressor $\varphi \left( t \right) \in {R^{n \times n}}$ is introduced.
\begin{assumption}
There exists the decomposition \eqref{1.6} with the time-invariant matrix  ${V_2}\left( t \right) \equiv {V_2}$ of $\varphi \left( t \right) \in {R^{n \times n}}$ with constant rank $r\left( t \right) \equiv r < n{\rm{,\;}}\overline r\left( t \right) \equiv \overline r > 0$.
\end{assumption}

Under Assumption 2, the disturbance $d\left( t \right) \equiv d$ and the unknown parameters $\Theta \left( t \right) \equiv \Theta$ are also time-invariant.

When the law \eqref{3.5} is applied and Assumption 2 is met, taking into account the results of Proposition 1, the properties of $\tilde z\left( t \right)$ and $\tilde \Theta \left( t \right)$ are analyzed in Theorem 2. In its first statement the unconditional properties are presented, in the second one the properties are shown that are guaranteed when the convergence condition is met, and in the third and fourth statements the asymptotic and exponential stability conditions are presented.

\begin{theorem} \label{th:2}
When Assumptions 1 and 2 are met, the following holds:
\begin{enumerate}
 \setlength{\leftskip}{15pt}
    \item[\rm{I.}]$\forall t \ge {t_0}{\rm{\;}}\left| {{{\tilde \Theta }_i}\left( {{t_a}} \right)} \right| \le \left| {{{\tilde \Theta }_i}\left( {{t_b}} \right)} \right|{\rm{\;}}\forall {t_a} \ge {t_b}.$
    \item[\rm{II.}]$\overline \varphi \left( t \right) \in {\rm{s \text{-} FE}} \Rightarrow \left\{ \begin{array}{l}
\left\| {\tilde \Theta \left( {{t_e}} \right)} \right\| \le \beta \left\| {\tilde \Theta \left( {t_r^ + } \right)} \right\|{\rm{;}}\\
\left| {\tilde z\left( {{t_e}} \right)} \right| \le \beta \left| {\tilde z\left( {t_r^ + } \right)} \right|.
\end{array} \right.$
    \item[\rm{III.}]$\omega \left( t \right) \notin {L_2} \Rightarrow \left\{ \begin{array}{l}
{\rm{li}}{{\rm{m}}_{t \to \infty }}\left\| {\tilde \Theta \left( t \right)} \right\| = 0{\rm{;}}\\
{\rm{li}}{{\rm{m}}_{t \to \infty }}\left| {\tilde z\left( t \right)} \right| = 0.
\end{array} \right.$
    \item[\rm{IV.}]$\overline \varphi \left( t \right) \in {\rm{s \text{-} PE}} \Rightarrow \left\{ \begin{array}{l}
{\rm{li}}{{\rm{m}}_{t \to \infty }}\left\| {\tilde \Theta \left( t \right)} \right\| = 0{\rm{ }}\left( {{\rm{exp}}} \right){\rm{;}}\\
{\rm{li}}{{\rm{m}}_{t \to \infty }}\left| {\tilde z\left( t \right)} \right| = 0{\rm{ }}\left( {{\rm{exp}}} \right).
\end{array} \right.$
\end{enumerate}
 
In this case the rate of exponential convergence can be directly adjusted by value of the parameter ${\gamma _0}$.
\end{theorem}

Proof of Theorem 2 is given in Appendix.

~

\begin{rem}
The asymptotic stability condition $\omega \left( t \right) \notin {L_2}$ is strictly weaker than the exponential one $\overline \varphi \left( t \right) \in {\rm{s \text{-} PE}}$, as, for example, there exists the regressor $\omega \left( t \right) = {\varepsilon ^{n - 1}}{\lambda _1}\left( t \right){\rm{,\;}}{\lambda _1}\left( t \right) = \frac{1}{{\sqrt {1 + t} }}{\rm{,}}$ such that $\omega \left( t \right) \notin {L_2}$ and $\overline \varphi \left( t \right) \notin {\rm{s \text{-} PE}}$ because \linebreak $\cancel{\exists }\mu  > 0{\rm{\;}}\forall t \ge {t_0}{\rm{\;}}{\lambda _1}\left( t \right) > \mu $, which contradicts Corollary  3. Therefore, when Assumption 2 is met, the weakest convergence condition of the law \eqref{3.5} to ensure convergence of the errors $\tilde \Theta \left( t \right)$, $\tilde z\left( t \right)$ to zero, and $\tilde \theta \left( t \right)$ to the set ${\theta _{{\rm{max}}}}$ is the non-square integrability of the multiplication of $r$ eigenvalues of $\varphi \left( t \right)$.
\end{rem}

\subsection{Piecewise-constant rank and nullspace basis}

The requirements of Assumption 2 are restrictive, and, as far as practical scenarios are concerned, both the rank and nullspace basis of the regressor usually change their values in piecewise-constant manner. Therefore, the properties of the law \eqref{3.5} are analyzed under the assumption that the rank and nullspace basis of $\varphi \left( t \right)$ are defined as piecewise-constant functions.

\begin{assumption}
The rank of $\varphi \left( t \right)$ is a piecewise-constant function, and there exists its decomposition \eqref{1.6} with the piecewise-constant matrix ${V_2}\left( t \right)$:
\begin{gather} \label{3.6}
\forall t \ge {t_0}{\rm{\;}}r\left( t \right) = \sum\limits_{{j_r} = 1}^\infty  {{\Delta _{{j_r}}}h\left( {t - {t_{{j_r}}}} \right)} {\rm{,\;}}{V_2}\left( t \right) = \sum\limits_{{j_V} = 1}^\infty  {{\Delta _{{j_V}}}h\left( {t - {t_{{j_V}}}} \right)} {\rm{,}}
\end{gather}
where ${t_{{j_r}}}$ is a time instant of rank change, ${\Delta _{{j_r}}}$ denotes the amplitude of rank change at time instant ${t_{{j_r}}}$,  ${t_{{j_V}}}$ stands for the time instant of change of the nullspace basis ${V_2}\left( t \right)$, \linebreak ${\Delta _{{j_V}}} \in {R^{n \times \overline r\left( t \right)}}$ is the amplitude of ${V_2}\left( t \right)$ change, $h\left( {t - {t_{{j_r}}}} \right){\rm{,\;}}h\left( {t - {t_{{j_V}}}} \right)$ are unit step functions.
\end{assumption}

When \eqref{3.6} is met, the disturbance $d\left( t \right)$ and unknown parameters $\Theta \left( t \right)$ are piecewise-constant functions: 
\begin{gather} \label{3.7}
d\left( t \right) = \sum\limits_{j = 1}^\infty  {{\Delta _j}h\left( {t - {t_j}} \right)} {\rm{,\;}}\dot d\left( t \right) = \sum\limits_{j = 1}^\infty  {{\Delta _j}\delta \left( {t - {t_j}} \right)} {\rm{,\;}}\Theta \left( t \right) = \theta  - \sum\limits_{j = 1}^\infty  {{\Delta _j}h\left( {t - {t_j}} \right)} {\rm{,}}
\end{gather}
where ${t_j} \in \left\{ {{t_{{j_r}}}{\rm{,\;}}{t_{{j_V}}}\left| {{j_r} \in \mathbb{N}{\rm{,\;}}{j_V} \in \mathbb{N}} \right.} \right\}$ are time instants of $d\left( t \right)$ change, $\delta \left( {t - {t_j}} \right)$ is a Dirac function, $\left\| {{\Delta _j}} \right\| \le {\Delta _{{\rm{max}}}}$ is a bounded value of the disturbance amplitude change.

Taking into consideration proved Proposition 1, the properties ensured by the law \eqref{3.5} when Assumptions 1 and 3 are met are stated in the following theorem.
\begin{theorem} \label{th:3}
Let the premises of Assumptions 1 and 3 hold and $\overline \varphi \left( t \right) \in {\rm{s \text{-} PE}}$ with the rank $ r\left( t \right) \ge 1$, then:
\begin{gather} \label{3.8}
\forall t \ge kT{\rm{\;}}\left\{ {\begin{array}{*{20}{c}}
{\left\| {\tilde \Theta \left( t \right)} \right\| \le a\left( {{t_j}} \right){e^{ - {\gamma _0}\left( {t - kT} \right)}}\left\| {\tilde \Theta \left( {kT} \right)} \right\|,}\\
{\left| {\tilde z\left( t \right)} \right| \le a\left( {{t_j}} \right){e^{ - {\gamma _0}\left( {t - kT} \right)}}\left| {\tilde z\left( {kT} \right)} \right|,}
\end{array}} \right.
\end{gather}
where $\left\{ {a\left( {{t_0}} \right){\rm{,\;}}a\left( {{t_1}} \right){\rm{,\;}} \ldots, a\left( {{t_j}} \right){\rm{,}} \ldots } \right\}$ is a numerical sequence.

Moreover, when  $\exists {a_{{\rm{max}}}}{\rm{\;}}\forall {t_j} \ge {t_0}{\rm{\;}}a\left( {{t_j}} \right) \le {a_{{\rm{max}}}}{\rm{,}}$ then $\tilde \Theta \left( t \right)$ and $\tilde z\left( t \right)$ are exponentially stable:
\begin{gather*}
\left\{ \begin{array}{l}
{\rm{li}}{{\rm{m}}_{t \to \infty }}\left| {\tilde z\left( t \right)} \right| = 0{\rm{\;}}\left( {{\rm{exp}}} \right){\rm{ }}\\
{\rm{li}}{{\rm{m}}_{t \to \infty }}\left\| {\tilde \Theta \left( t \right)} \right\| = 0{\rm{\;}}\left( {{\rm{exp}}} \right)
\end{array} \right..
\end{gather*}
\end{theorem}

Proof of Theorem 3 and the definition of $a\left( {{t_j}} \right)$ are presented in Appendix.

~

On the one hand, the results of Theorem 3 show the robustness of the law \eqref{3.5} to variations of the rank and nullspace basis of the regressor $\varphi \left( t \right)$ in the sense of exponential recovery of equilibrium points of the errors $\tilde \Theta \left( t \right)$ and $\tilde z\left( t \right)$, and on the other hand, describe necessary and sufficient conditions of such errors exponential convergence to zero. These conditions are the regressor semi-persistent excitation with rank not less than one and the fact that the inequalities $a\left( {{t_j}} \right) \le {a_{{\rm{max}}}}$ hold for all ${t_j} \ge {t_0}$.

However, Theorem 3 does not provide a constructive description of the requirements for $a\left( {{t_j}} \right)$ or ${\Delta _j}$, which, being met for all ${t_j} \ge {t_0}$, guarantee $a\left( {{t_j}} \right) \le {a_{{\rm{max}}}}$ and hence exponential stability of the errors $\tilde \Theta \left( t \right)$ and $\tilde z\left( t \right)$ when the rank or nullspace basis are piecewise-constant functions.

In the following corollary, we introduce two additional conditions, under which for all ${t_j} \ge {t_0}$ it is ensured that the inequality $a\left( {{t_j}} \right) \le {a_{{\rm{max}}}}$ holds.
\begin{corollary} \label{cor5} 
Let the premises of Theorem 3 be met and additionally one of the following conditions also hold: 
\begin{enumerate}
\setlength{\leftskip}{15pt}
\item[$1)$]$j \le {j_{{\rm{max}}}} < \infty {\rm{;}}$
\item[$2)$]${\Delta _{{\rm{max}}}} \le c\left( {{t_j}} \right){e^{ - {\gamma _0}\left( {{t_j} - kT} \right)}}{\rm{,\;}}\forall j \in \mathbb{N}{\rm{\;}}c\left( {{t_j}} \right) \ge c\left( {{t_{j + 1}}} \right) > 0.$
\end{enumerate}

Then there exists ${a_{{\rm{max}}}}$ such that $\forall {t_j} \ge {t_0}{\rm{\;}}a\left( {{t_j}} \right) \le {a_{{\rm{max}}}}$.
\end{corollary}

Proof of Corollary 5 is given in Appendix.

~

According to the results of Corollary 5, the condition $a\left( {{t_j}} \right) \le a_{{{\rm{max}}}}$ is met when the norm of the parameter change value $\Delta _{{{\rm{max}}}}$ is upper bounded by a decreasing sequence, or when a number of regressor nullspace base/rank switches $j$ is finite.

\subsection{Conditions of partial identifiability}

Considering the identification problems, the main aim is to ensure the convergence of the parameter error $\tilde \theta \left( t \right)$. Therefore, in addition to the results of Sections 3.2 and 3.3, the conditions are defined under which the elements of the vector of new unknown parameters $\Theta \left( t \right)$ partially or completely coincide with the elements of the original vector $\theta$.

The analysis of the parameters $\Theta \left( t \right)$ properties are written as a proposition.
\begin{proposition}\label{st2}
Let the matrix $\overline \Lambda \left( t \right)$ be obtained with the help of (3.2) when $\overline \varepsilon  = 0$, then:
\begin{enumerate}
\setlength{\leftskip}{15pt}
\item[$1)$]${{\overline \varphi \left( t \right) \in {\rm{FE}}}/{\overline \varphi \left( t \right) \in {\rm{PE}}}}\Rightarrow \Theta \left( t \right) = \theta {\rm{;}}$
\item[$2)$] if Assumption 2 and the following conditions are met:
\begin{gather*}
\begin{array}{c}
{\overline \varphi \left( t \right) \in {\rm{s\text{-} FE}}}/ {\overline \varphi \left( t \right) \in {\rm{s\text{-}{PE}}}}{\rm{,\;}}n > 2,\\
\sum\limits_{i = 1}^{n - p} {{w_i}{\varphi _i}\left( t \right)}  + \sum\limits_{j = n - p + 1}^n {{w_j}{\varphi _j}\left( t \right)}  = {0_n}{\rm{,\;}}{w_i} \ne 0,{\rm{\;}}{w_j} = 0,
\end{array}
\end{gather*}
\end{enumerate}
then $\exists M \subset \left\{ {1,...,n} \right\}{\rm{,\;}}\left| M \right| = p{\rm{,\;}}\forall i \in M{\rm{,\;}}{\Theta _i} = {\theta _i}.$
\end{proposition}

Proof of Proposition 2 is presented in Appendix\footnote{In statement (2) of Proposition 2, without loss of generality, it is assumed that the first $n-p$ columns of the regressor $\varphi \left( t \right) = \left[ {{\varphi _1}\left( t \right) \ldots {\varphi _i}\left( t \right) \ldots {\varphi _n}\left( t \right)} \right]$ are linearly dependent (in case $\overline r\left( t \right) > 0$ such form can always be obtained by columns permutation).}.

~

Thus, according to Proposition 2, the conditions of partial identifiability of parameters $\theta$ are: (1) Assumption 2 is met, (2) the regressor $\overline \varphi \left( t \right)$ is semi-persistently exciting, (3) $p$ columns of the regressor $\varphi \left( t \right)$ are linearly independent, (4) the identification problem dimension is $n > 2$. Combining the results of Theorem 1 and Proposition 2, a corollary is obtained that describes the convergence conditions for a part of parameter errors ${\tilde \theta _i}\left( t \right)$.

\begin{corollary} 
Let Assumptions 1-2 and the following conditions be met: 
\begin{gather*}
\sum\limits_{i = 1}^{n - p} {{w_i}{\varphi _i}\left( t \right)}  + \sum\limits_{j = n - p + 1}^n {{w_j}{\varphi _j}\left( t \right)}  = {0_n}{\rm{,\;}}{w_i} \ne 0,{\rm{\;}}{w_j} = 0,{\rm{\;}}n > 2.
\end{gather*}
Then:
\begin{enumerate}
\setlength{\leftskip}{15pt}
\item[$a)$] $\overline \varphi \left( t \right) \in {\rm{s \text{-} FE}} \Leftrightarrow \forall i \in M{\rm{\;}}\left\{ \begin{array}{l}
\left| {{{\tilde \theta }_i}\left( {{t_e}} \right)} \right| \le \beta \left| {{{\tilde \theta }_i}\left( {t_r^ + } \right)} \right|\\
\left| {{{\tilde \theta }_i}\left( {{t_a}} \right)} \right| \le \left| {{{\tilde \theta }_i}\left( {{t_b}} \right)} \right|{\rm{\;}}\forall {t_a} \ge {t_b}
\end{array} \right.;$
\item[$b)$]$\omega \left( t \right) \notin {L_2} \Leftrightarrow \forall i \in M{\rm{\;}}\left\{ \begin{array}{l}
{\rm{li}}{{\rm{m}}_{t \to \infty }}\left| {{{\tilde \theta }_i}\left( t \right)} \right| = 0\\
\left| {{{\tilde \theta }_i}\left( {{t_a}} \right)} \right| \le \left| {{{\tilde \theta }_i}\left( {{t_b}} \right)} \right|{\rm{\;}}\forall {t_a} \ge {t_b}
\end{array} \right.;$
\item[$c)$]$\overline \varphi \left( t \right) \in {\rm{s \text{-} PE}} \Leftrightarrow \forall i \in M{\rm{\;}}\left\{ \begin{array}{l}
{\rm{li}}{{\rm{m}}_{t \to \infty }}\left| {{{\tilde \theta }_i}\left( t \right)} \right| = 0{\rm{\;}}\left( {{\rm{exp}}} \right)\\
\left| {{{\tilde \theta }_i}\left( {{t_a}} \right)} \right| \le \left| {{{\tilde \theta }_i}\left( {{t_b}} \right)} \right|{\rm{\;}}\forall {t_a} \ge {t_b}.
\end{array} \right.$
\end{enumerate}
\end{corollary}

Corollary 6 is obtained by combining the consistent premises and results of \linebreak Theorem 1 and Statement 2.

~

\begin{rem}
It is worth noting the existence of regressors $\varphi \left( t \right)$ that do not satisfy the requirements of Proposition 2, but still ensure the existence of zero elements in the vector $d$ and allow one to identify some of the original unknown parameters $\theta$. For such regressors, the fact that some elements of $d$ are zero is not caused by the existence of zero rows/columns in the product $V_2^{\rm{T}}{V_2}$ (see the proof of Proposition 2), but by the equality to zero of the elements of the product $V_2^{\rm{T}}{V_2}\theta$ (due to orthogonality of ${V_2}$ and $\theta$).

For an instance, if $\varphi \left( t \right) = \left[ {\begin{array}{*{20}{c}}
1&{ - 1}\\
{ - 1}&1
\end{array}} \right]{\rm{,\;}}\theta  = \vartheta \left[ {\begin{array}{*{20}{c}}
{ - 1}&1
\end{array}} \right]{\rm{,\;}}\vartheta  \ne 0$, then the premises of Proposition 2 do not hold, but $d = {0_n}{\rm{,}}\;\Theta  = \theta$.
\end{rem}

\begin{rem}
From the practical point of view, it is important not only to prove that some elements of the parameter vector $\Theta$ coincide with the elements of $\theta$ under some conditions, but also to indicate their positions in such vector. For this purpose, the indices of the zero rows of the basis ${V_2}$ can be used as such indicators if the premises of statement 2 of Proposition 2 are satisfied.
\end{rem}

\begin{rem}
Under Assumption 3, the results of statement 2 of Proposition 2 are true locally over the time intervals when the regressor rank and nullspace basis are time-invariant. Hence, when the rank $r\left( t \right)$ changes its value, different number $p$ of elements of the vector $\theta$ can be identified over different time ranges $\left[ {{t_{j - 1}}{\rm{;\;}}{t_j}} \right]$ and $\left[ {{t_j}{\rm{;\; }}{t_{j + 1}}} \right]$, and when the regressor nullspace basis changes its value,  different elements of vector $\theta$ can be identified over different time intervals $\left[ {{t_{j - 1}}{\rm{; \;}}{t_j}} \right]$ and $\left[ {{t_j}{\rm{;\;}}{t_{j + 1}}} \right]$.
\end{rem}

\setcounter{equation}{0}  
\renewcommand{\theequation}{\arabic{section}.\arabic{subsection}.\arabic{equation}}

\section{Mathematical Modelling}
The DREM identification law with regularization~\eqref{3.5} has been compared with the classical gradient~\eqref{2.3} and DREM without regularization~\eqref{2.7} ones in Matlab/Simulink. The simulation was conducted using numerical integration by the Euler method with a fixed discretization step ${\tau _s} = {10^{ - 4}}$ second.

Sections 4.1 and 4.2 presents the obtained simulation results under Assumptions 2 and 3 respectively.

\subsection{Time-invariant rank and nullspace basis of regressor}
The regression equation~\eqref{2.1} was defined as:
\begin{gather} \label{4.1.1}
z\left( t \right) = {\overline \varphi ^{\text{T}}}\left( t \right)\theta  = \left[ {\begin{array}{*{20}{c}}
{ - 2{e^{ - t}}\cos \left( t \right)}&{{e^{ - t}}\cos \left( t \right)}&{{e^{ - t}}} 
\end{array}} \right]\left[ {\begin{array}{*{20}{c}}
  4 \\ 
  { - 8} \\ 
  {12} 
\end{array}} \right].
\end{gather}

The parameters of the filter~\eqref{2.4}, algorithm of the eigenvalue virtual substitution~\eqref{3.2} and identification laws~\eqref{3.5},~\eqref{2.3} were set as:
\begin{gather} \label{4.1.2}
l = 100,{\text{ }}\varepsilon  = 0,4,{\text{ }}\overline \varepsilon  = {10^{ - 10}}{\text{, }}{\gamma _0} = 5,{\text{ }}{\gamma _1} = 1,{\text{ }}\Gamma  = 5{I_3}.
\end{gather}

In order to provide the same convergence rate for the laws~\eqref{3.5} and~\eqref{2.7}, the adaptive gain $\gamma$ of the law~\eqref{2.7} was defined similarly to~\eqref{3.5}, following the method of the regressor excitation normalization \cite{18}:
\begin{gather} \label{4.1.3}
\gamma \left( t \right) = \left\{ \begin{gathered}
  {\gamma _1}{\text{, if }}\omega \left( t \right) \leqslant {\text{min}}\left\{ {\lambda _{{\text{min}}}^n\left( t \right){\text{, }}{\varepsilon ^n}} \right\} \hfill \\
  \tfrac{{{\gamma _0}}}{{{\omega ^2}\left( t \right)}}{\text{ otherwise}} \hfill \\ 
\end{gathered}  \right..
\end{gather}

First of all, it was shown that the convergence conditions of the laws~\eqref{2.3},~\eqref{2.7} and~\eqref{3.5} were met. Figure 1 presents the behaviour of the disturbance $d$ and the rank of regressor $\varphi \left( t \right)$ in the course of the experiment.

\begin{figure}[h!]
\begin{minipage}[h]{0.49\linewidth}
\center{\includegraphics[width=1\linewidth]{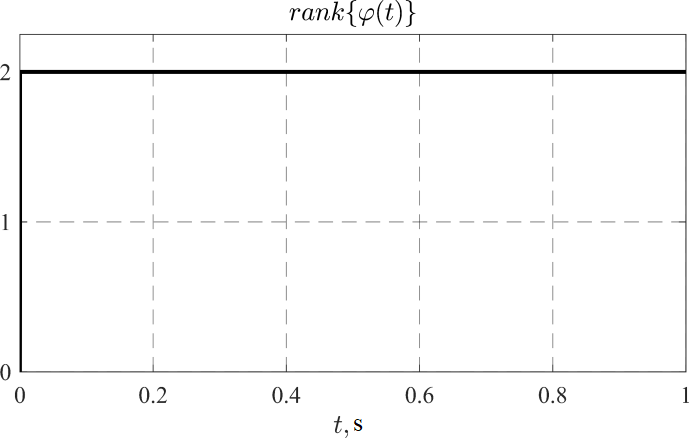} \\ a)}
\end{minipage}
\hfill
\begin{minipage}[h]{0.49\linewidth}
\center{\includegraphics[width=1\linewidth]{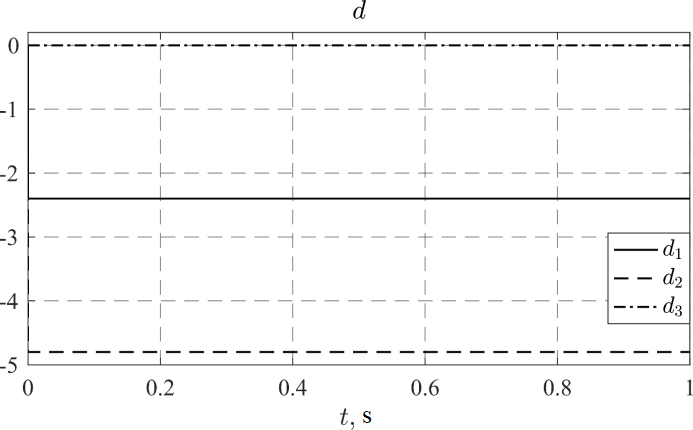} \\ b)}
\end{minipage}
\caption{Rank of the regressor $\varphi \left( t \right)$ (a), the disturbance $d$ (b).}
\label{ris:image1}
\end{figure}

As follows from the definition of the regressor $\overline \varphi \left( t \right)$, Fig. 1,a, the convergence conditions $\left( \overline \varphi \left( t \right) \in {\rm{s \text{-} FE}} \right)$ of laws~\eqref{2.3} and~\eqref{3.5} were met for all $t \geqslant 0$, whereas the convergence condition $\left( {\overline \varphi \left( t \right) \in {\text{FE}}} \right)$ of the law~\eqref{2.7} was not satisfied, so the simulation results are given only for the algorithms~\eqref{3.5} and~\eqref{2.3}. It followed from Fig. 1,a,b, that Assumption 2 was met, and, consequently, since ${\overline \varphi \left( t \right) \in \rm{s \text{{-}FE}}}$, the law~\eqref{3.5} guaranteed the errors $\tilde \Theta \left( t \right),\;\tilde z\left( t \right)$ reduction in the course of the experiment. Moreover, as Assumption 2 was satisfied, ${d_3} = 0$ and $r = 2$, then the law~\eqref{3.5} additionally ensured that the error ${\tilde \theta _3}\left( t \right)$ decreased.

Firstly, it was set that ${\theta _0} = {{\begin{bmatrix}
  0&0&0 
\end{bmatrix}}^{\text{\rm T}}}$, which meant that, according to Theorem 1, the law~\eqref{3.5} was quasi-convergent (the reduction of $\left| {\tilde z\left( t \right)} \right|$ was guaranteed, as well as the lack of growth of $\left\| {\tilde \theta \left( t \right)} \right\|$ over the time range $\left[ {0{\text{; 1}}} \right]$).

Figure 2 presents the transients of errors ${\tilde \theta _i}\left( t \right)$ of the laws~\eqref{3.5} – (a) and~\eqref{2.3} – (b).

\begin{figure}[h!]
\begin{minipage}[h]{0.49\linewidth}
\center{\includegraphics[width=1\linewidth]{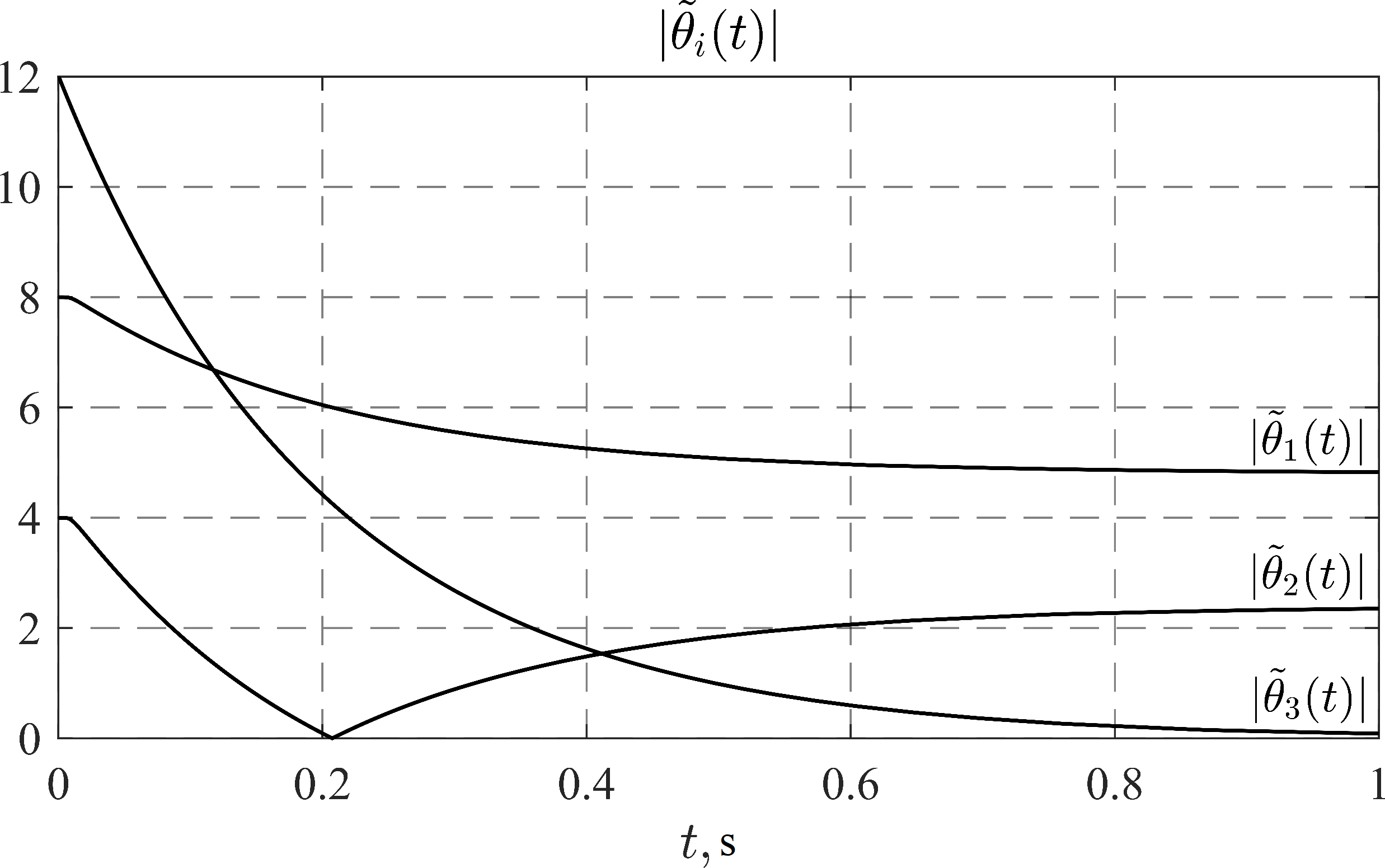} \\ a)}
\end{minipage}
\hfill
\begin{minipage}[h]{0.49\linewidth}
\center{\includegraphics[width=1\linewidth]{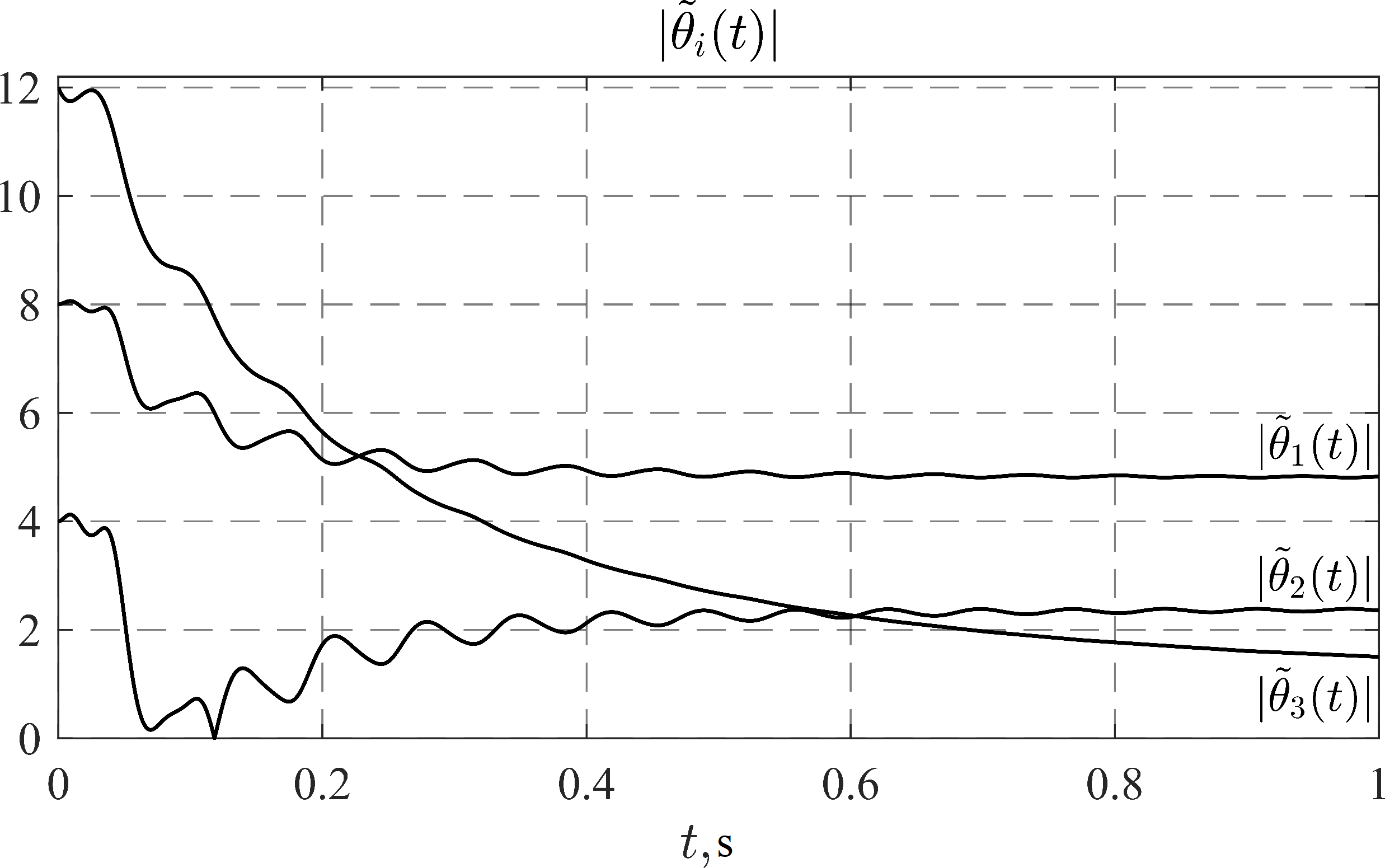} \\ b)}
\end{minipage}
\caption{Transient curves of the errors ${\tilde \theta _i}\left( t \right)$ of the laws~\eqref{3.5} – (a) and~\eqref{2.3} – (b).}
\label{ris:image2}
\end{figure}

The obtained transients indicate the advantages of~\eqref{3.5} over~\eqref{2.7} and the classical gradient~\eqref{2.3} identification laws. In particular, unlike~\eqref{2.7}, the law~\eqref{3.5} reduced the {\it a priori} values of the errors ${\tilde \theta _i}\left( t \right)$ and, unlike~\eqref{2.3}, ensured the  transients of first-order type and monotonic exponential convergence of the error ${\tilde \theta _3}\left( t \right)$ to zero. The monotonicity of ${\tilde \theta _1}\left( t \right)$ can be explained by the fact that the condition ${\theta _1} \leqslant {\Theta _1}{\text{, }}{\hat \theta _1}\left( {{t_0}} \right) > {\Theta _1}$ was met in the course of the experiment, which was a particular favorable situation.

Figure 3,a shows a comparison of the error $\tilde z\left( t \right)$ curves of the laws~\eqref{3.5} and~\eqref{2.3}, while Figure 3,b presents the transients of the error ${\tilde \Theta _i}\left( t \right)$ when the law~\eqref{3.5} was applied.

\begin{figure}[h!]
\begin{minipage}[h]{0.49\linewidth}
\center{\includegraphics[width=1\linewidth]{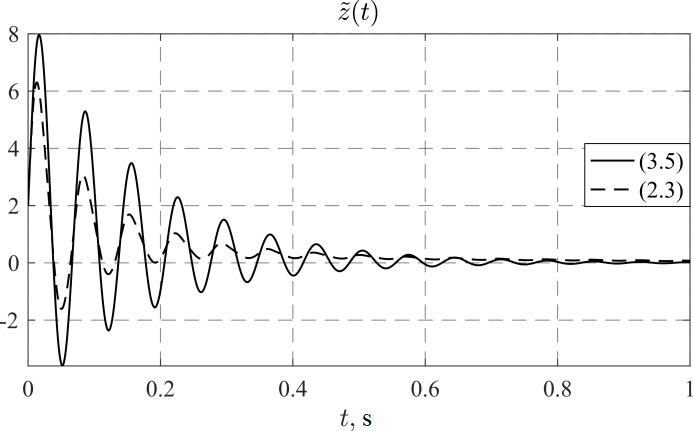} \\ a)}
\end{minipage}
\hfill
\begin{minipage}[h]{0.49\linewidth}
\center{\includegraphics[width=1\linewidth]{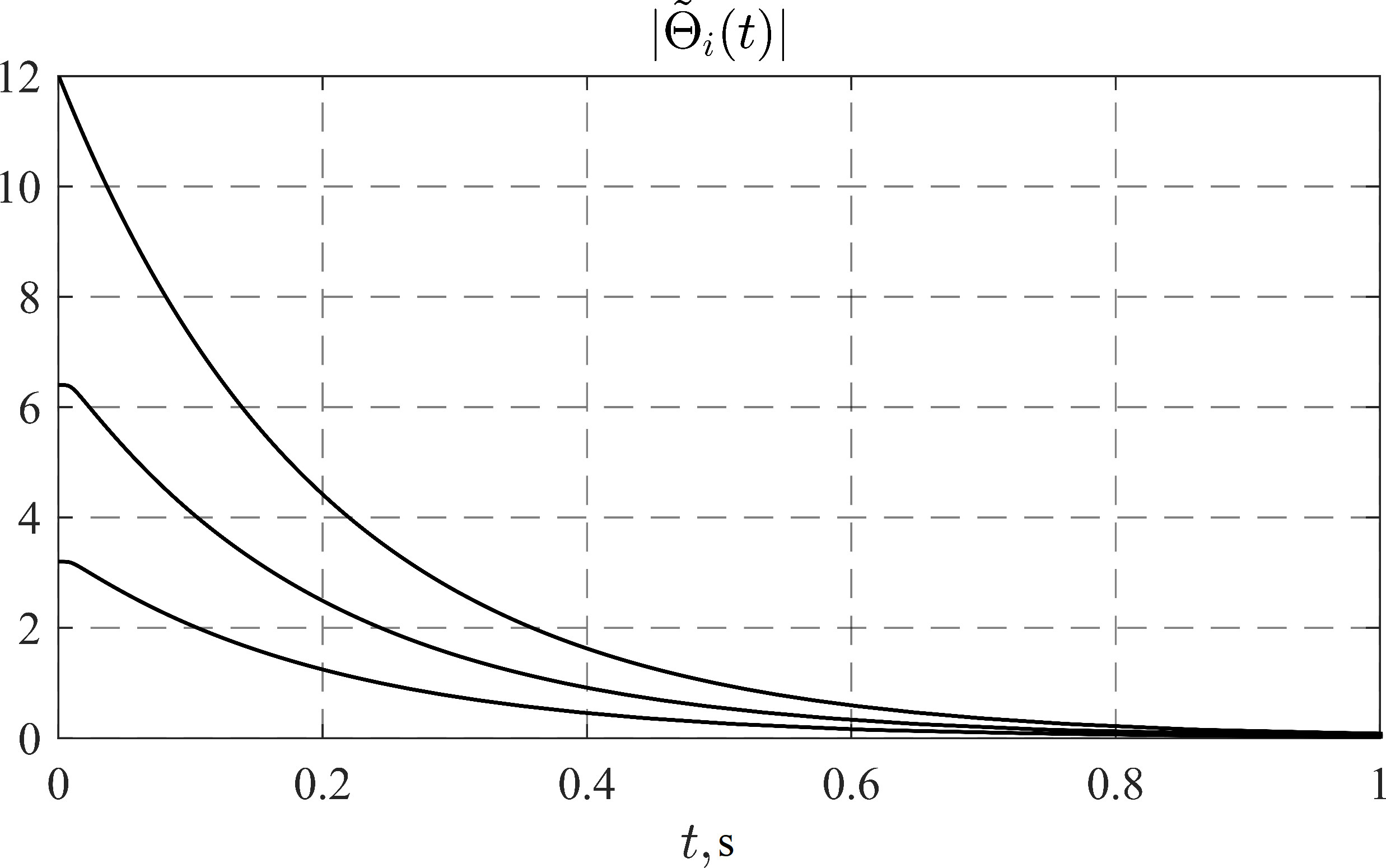} \\ b)}
\end{minipage}
\caption{Transient curves of (a) the error $\tilde z\left( t \right)$ of the laws~\eqref{3.5} and~\eqref{2.3} and (b) the errors ${\tilde \Theta _i}\left( t \right)$ of the law~\eqref{3.5}.}
\label{ris:image3}
\end{figure}

Figure 3,a confirms that $\tilde z\left( t \right)$ was reduced over the time range $\left[ {0{\text{; 1}}} \right]$ when the law~\eqref{3.5} was applied, Figure 3, b demonstrates the monotonicity of the error ${\tilde \Theta _i}\left( t \right){\text{ }}\forall i \in \overline {1,n} $, which was proved analytically in Theorem 2.

Figure 4 shows the behaviour of $\left\| {\tilde \theta \left( t \right)} \right\|$ obtained with the help of the laws~\eqref{3.5} and~\eqref{2.3} under different initial conditions (for all initial conditions the law~\eqref{3.5} was convergent or quasi-convergent).

\begin{figure}[h!]
\begin{minipage}[h]{0.49\linewidth}
\center{\includegraphics[width=1\linewidth]{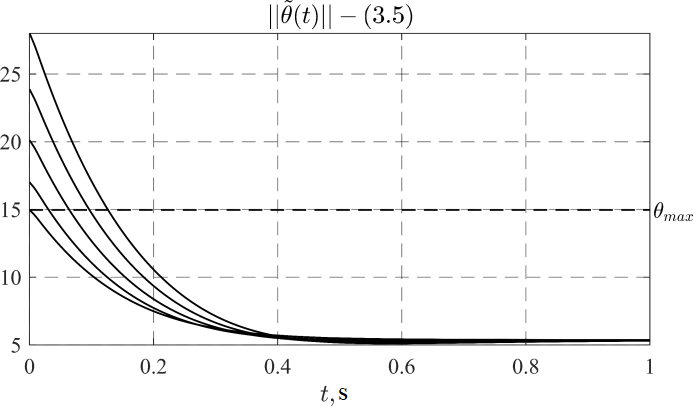} \\ a)}
\end{minipage}
\hfill
\begin{minipage}[h]{0.49\linewidth}
\center{\includegraphics[width=1\linewidth]{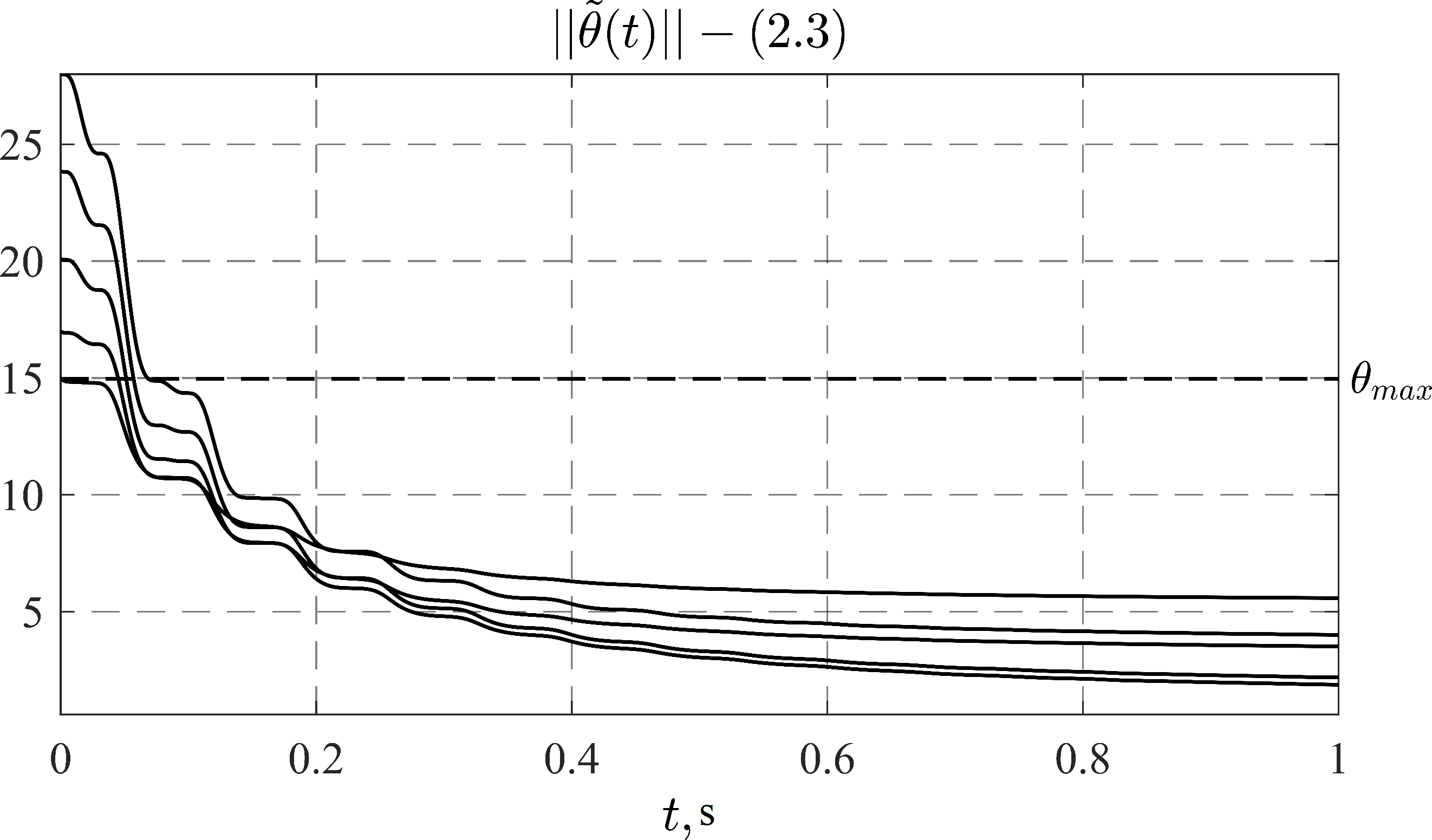} \\ b)}
\end{minipage}
\caption{Transient curves of $\left\| {\tilde \theta \left( t \right)} \right\|$ under different initial conditions.}
\label{ris:image4}
\end{figure}

The transients in Fig. 4 confirm the exponential convergence of the error $\tilde \theta \left( t \right)$ to a set with the bound $\theta _{{{\text{max}}}}$ proved in Theorem 1.

Then it was set that ${\theta _0} = {{\begin{bmatrix}
  0&{ - 10}&{14} 
\end{bmatrix}}^{\text{\rm T}}}$, which did not satisfy the convergence conditions from Theorem 1 since $\left\| {\tilde \theta \left( {t_r^ + } \right)} \right\| \approx 4.9$ and \linebreak ${\theta _{{\text{max}}}} = \left\| \theta \right\| \approx $15. Figure 5 shows the behaviour of $\left\| {\tilde \theta \left( t \right)}\right\|$ under such choice of the initial conditions when the laws~\eqref{3.5} and~\eqref{2.3} were used.

\begin{figure}[h!]
\center
\includegraphics[width=0.5\linewidth]{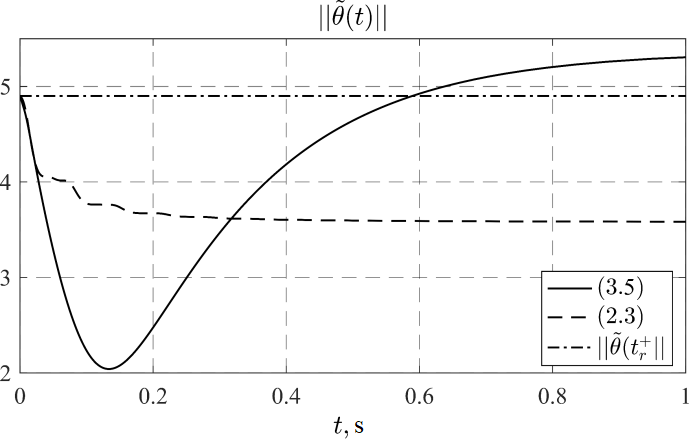}
\caption{Transient curves of $\left\| {\tilde \theta \left( t \right)} \right\|$ for the laws~\eqref{3.5} and~\eqref{2.3}.}
\label{ris:image5}
\end{figure}

The transients of $\left\| {\tilde \theta \left( t \right)} \right\|$ shown in Fig. 5 validated the conclusions made in Theorem 1. 
Indeed, when $\left\| {\tilde \theta \left( {t_r^ + } \right)} \right\| < {\theta _{{\text{max}}}}$, the convergence condition of the law~\eqref{3.5} was not met, and, consequently, the error norm $\left\| {\tilde \theta \left( t \right)} \right\|$ could become greater than $\left\| {\tilde \theta \left( {t_r^ + } \right)} \right\|$, and it was not ensured that all conditions of~\eqref{2.2} were met.

Thus, the conducted numerical experiments fully confirmed the properties of the law~\eqref{3.5} described within Theorems 1-2, Proposition 2 and Corollary 6 when $\overline \varphi \left( t \right) \in {\rm{s \text{-} FE}}$ and Assumption 2 was met.

\setcounter{equation}{0}  

\subsection{Piecewise-constant rank and nullspace basis of regressor}

\subsubsection{\text{First experiment}}

The regression equation~\eqref{2.1} was defined as follows:

\begin{gather} \label{4.2.1}
\begin{gathered}
  z\left( t \right) = {{\overline \varphi }^{\text{T}}}\left( t \right)\theta  = \left[ {\begin{array}{*{20}{c}}
  {{{\overline \varphi }_1}\left( t \right)}&{{{\overline \varphi }_2}\left( t \right)}&{{{\overline \varphi }_3}\left( t \right)} 
\end{array}} \right]\left[ {\begin{array}{*{20}{c}}
  4 \\ 
  { - 8} \\ 
  {12} 
\end{array}} \right]{\text{,}} \\ 
  {{\overline \varphi }_1}\left( t \right) = 9sin\left( t \right){\text{; }}{{\overline \varphi }_2}\left( t \right) = \left\{ \begin{gathered}
  2sin\left( t \right){\text{, 0}} \leqslant t \leqslant 5 \hfill \\
  {\text{4}}{\text{, 5}} < t \leqslant 15 \hfill \\
  2sin\left( t \right){\text{, }}t > 15 \hfill \\ 
\end{gathered}  \right.{\text{; }}\\{{\overline \varphi }_3}\left( t \right) = \left\{ \begin{gathered}
  sin\left( t \right){\text{, 0}} \leqslant t \leqslant 10 \hfill \\
  sin\left( {50t} \right){\text{, 10}} < t \leqslant 15 \hfill \\
  sin\left( t \right){\text{, }}t > 15 \hfill \\ 
\end{gathered}  \right.. \\ 
\end{gathered}
\end{gather}

The parameters of the filter~\eqref{2.4}, algorithm of the eigenvalue virtual substitution~\eqref{3.2} and identification laws~\eqref{2.3},~\eqref{3.5} were set as:
\begin{gather} \label{4.2.2}
\begin{gathered}
 l = 100,{\text{ }}\varepsilon  = 0,4,{\text{ }}\overline \varepsilon  = {10^{ - 10}}{\text{, }}{\gamma _0} = 5,\;{\gamma _1} = 1,{\text{ }}\Gamma  = {I_3}.
\end{gathered}
\end{gather}
	
In order to provide the same convergence rate for the laws~\eqref{3.5} and~\eqref{2.7}, the adaptive gain $\gamma$ of the law~\eqref{2.7} was defined similarly to~\eqref{3.5}, following the method of the regressor excitation normalization \cite{18}:
\begin{gather} \label{4.2.3}
\gamma \left( t \right) = \left\{ \begin{gathered}
  {\gamma _1}{\text{, if }}\omega \left( t \right) \leqslant {\text{min}}\left\{ {\lambda _{{\text{min}}}^n\left( t \right){\text{, }}{\varepsilon ^n}} \right\} \hfill \\
  \tfrac{{{\gamma _0}}}{{{\omega ^2}\left( t \right)}}{\text{ otherwise }} \hfill \\ \end{gathered}  \right..
\end{gather}

First of all, it was shown that the convergence conditions of the laws~\eqref{2.3},~\eqref{2.7} and~\eqref{3.5} were met. Figure 6 presents the behaviour of the disturbance $d\left( t \right)$ and rank of the regressor $\varphi \left( t \right)$ in the course of the experiment.

\begin{figure}[h!]
\begin{minipage}[h]{0.49\linewidth}
\center{\includegraphics[width=1\linewidth]{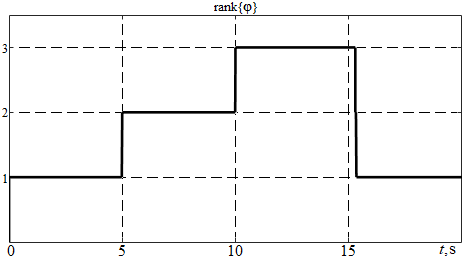} \\ a)}
\end{minipage}
\hfill
\begin{minipage}[h]{0.49\linewidth}
\center{\includegraphics[width=1\linewidth]{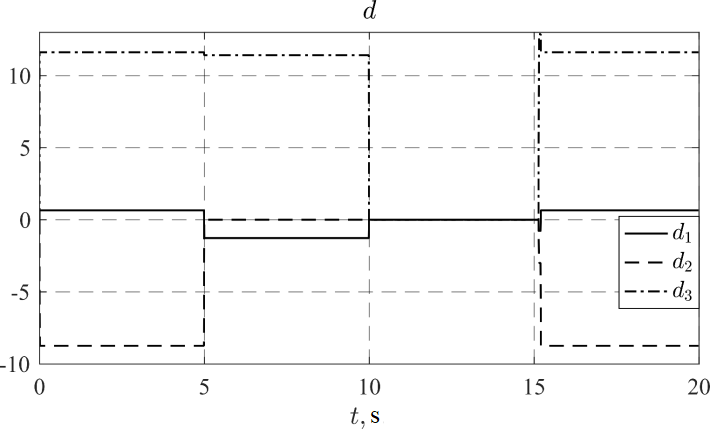} \\ b)}
\end{minipage}
\caption{Rank of the regressor $\varphi \left( t \right)$ (a), the disturbance $d\left( t \right)$ (b).}
\label{ris:image6}
\end{figure}

As follows from Figure 6,a and Corollaries 2 and 4, the necessary condition of the convergence of~\eqref{3.5} was met for all $t \geqslant 0$, while the convergence condition of~\eqref{2.7} was satisfied only over the time range $t \in \left[ {10{\text{; 15}}{\text{,34}}} \right]$. According to Fig. 6 the number of parameter switches was finite $j \leqslant {j_{{\text{max}}}} < \infty $ and $r\left( t \right) \geqslant 1$, and then, according to the results of Theorem 3 and Corollary 5, all necessary and sufficient conditions of exponential convergence of errors $\tilde z\left( t \right)$ and $\tilde \Theta \left( t \right)$ to zero were satisfied for~\eqref{3.5}. Moreover, since $\forall t \in \left[{5{\text{;10}}} \right] {\text{\;}}{d_2}\left( t \right) = 0$, the partial identifiability conditions from Proposition 2 were also met over the time range $\left[ {5{\text{;\;10}}} \right]$.

Having verified that the convergence conditions were met, the experiments were conducted using the algorithms~\eqref{3.5},~\eqref{2.7} and~\eqref{2.3} under different initial conditions.

Firstly, it was set that ${\theta _0} = {{\begin{bmatrix}
  0&5&0 
\end{bmatrix}}^{\text{\rm T}}}$, which, according to Theorem 1, ensured that the necessary conditions of convergence of the law~\eqref{3.5} were met:
\begin{displaymath}
{\beta _1} = \left\| {\tilde \theta \left( {t_r^ + } \right)} \right\|{\left\| \theta  \right\|^{ - 1}} \approx \tfrac{{18}}{{15}} = 1,2{\text{; }}\tfrac{1}{{{\beta _1}}} + {e^{ - {\gamma _0}\delta }} = \tfrac{1}{{1,2}} + {e^{ - 5 \cdot 5}} \approx 0,833 \in \left( {0{\text{; 1}}} \right).
\end{displaymath}

Figure 7 depicts the transients of ${\tilde \theta _i}\left( t \right)$ for~\eqref{3.5} – (a),\linebreak \eqref{2.7} – (b) and~\eqref{2.3} – (c).

\begin{figure}[h!]
\begin{minipage}[h]{0.49\linewidth}
\center{\includegraphics[width=1\linewidth]{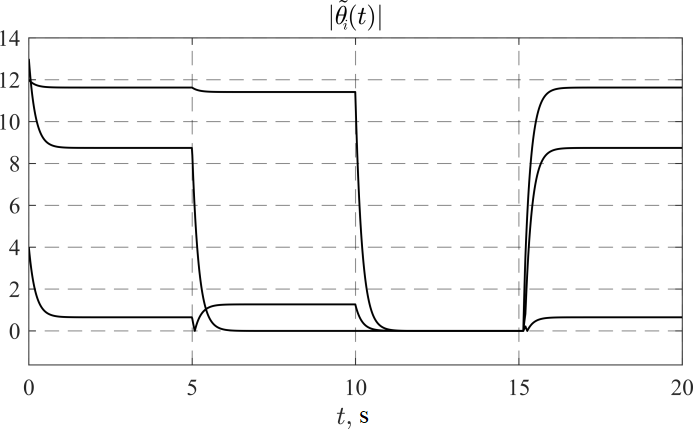} \\ a)}
\end{minipage}
\hfill
\begin{minipage}[h]{0.49\linewidth}
\center{\includegraphics[width=1\linewidth]{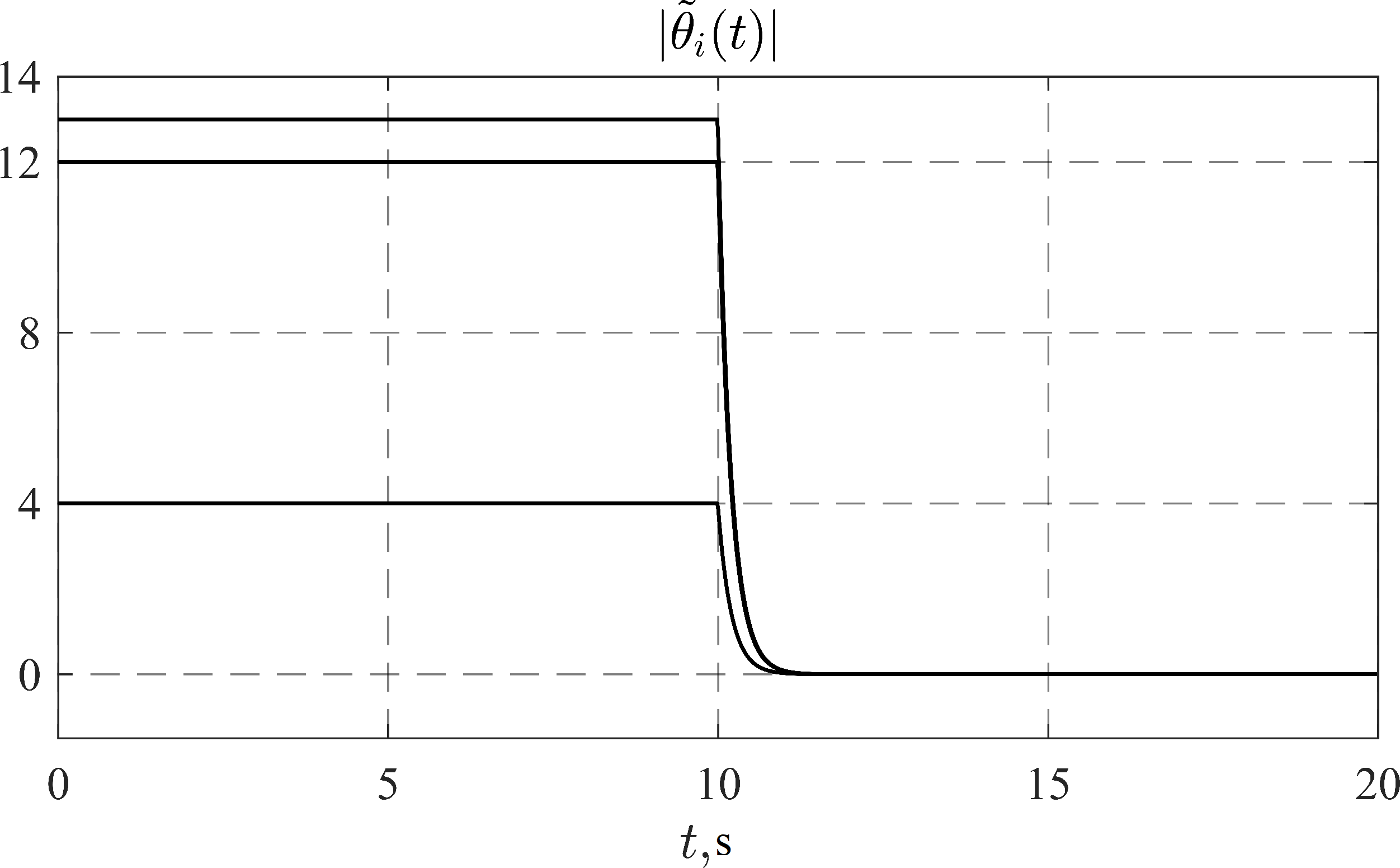} \\ b)}
\end{minipage}
\vfill
\center{\includegraphics[width=0.5\linewidth]{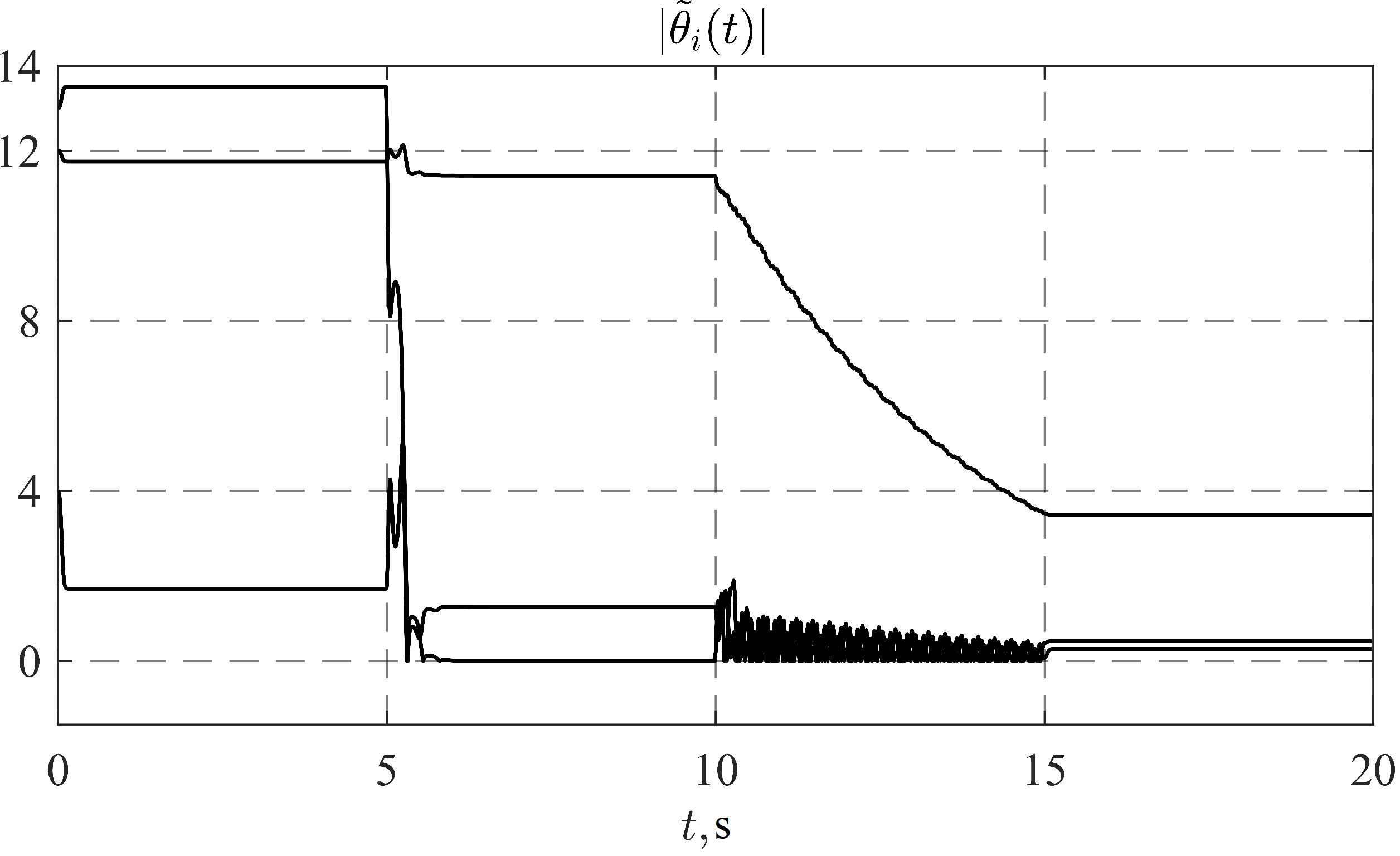} \\ c)}
\caption{Transient curves of ${\tilde \theta _i}\left( t \right)$ of the laws~\eqref{3.5} – (a),~\eqref{2.7} – (b) and~\eqref{2.3} – (c).}
\label{ris:image7}
\end{figure}

The obtained transients confirmed the theoretical conclusions made in Remark 4. Indeed, if the conditions of the second statement of Proposition 2 were met over $\left[ {5{\text{; 10}}} \right]$, then the law~\eqref{3.5}, in contrast to~\eqref{2.7} and~\eqref{2.3}, ensured monotonicity for one element of the vector $\tilde \theta \left( t \right)$. Comparing the quality of the transients, the advantages of the law~\eqref{3.5} over~\eqref{2.7} and~\eqref{2.3} are seen. As for~\eqref{2.3}, the law~\eqref{3.5} guaranteed the first-order type transient of ${\tilde \theta _i}\left( t \right){\text{\;}}\forall i \in \overline {1,n}.$ In comparison with~\eqref{2.7}, the law~\eqref{3.5} converged not only over the time range $\left[ {10{\text{; 15}}{\text{,34}}} \right]$, but for all $t \geqslant 0$, and ensured that one element of the vector $\tilde \theta \left( t \right)$ decreased to zero over $\left[ 5{\text{; 10}} \right]$.

Figure 8 presents the transients of $\tilde z\left( t \right)$ for the control systems based on the laws~\eqref{3.5} – (a),~\eqref{2.7} – (b) and~\eqref{2.3} – (c).

\begin{figure}[h!]
\begin{minipage}[h]{0.49\linewidth}
\center{\includegraphics[width=1\linewidth]{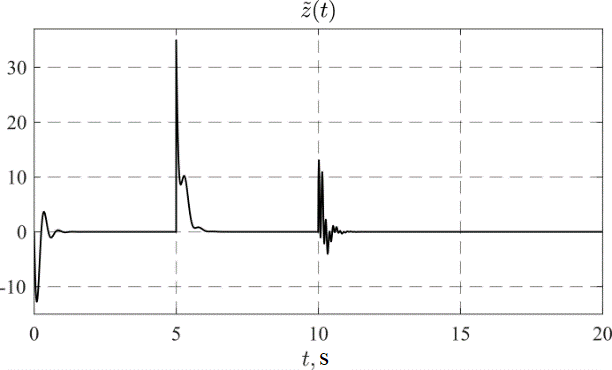} \\ a)}
\end{minipage}
\hfill
\begin{minipage}[h]{0.49\linewidth}
\center{\includegraphics[width=1\linewidth]{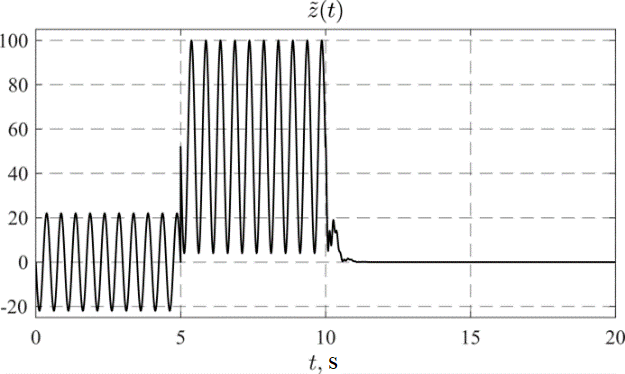} \\ b)}
\end{minipage}
\vfill
\center{\includegraphics[width=0.5\linewidth]{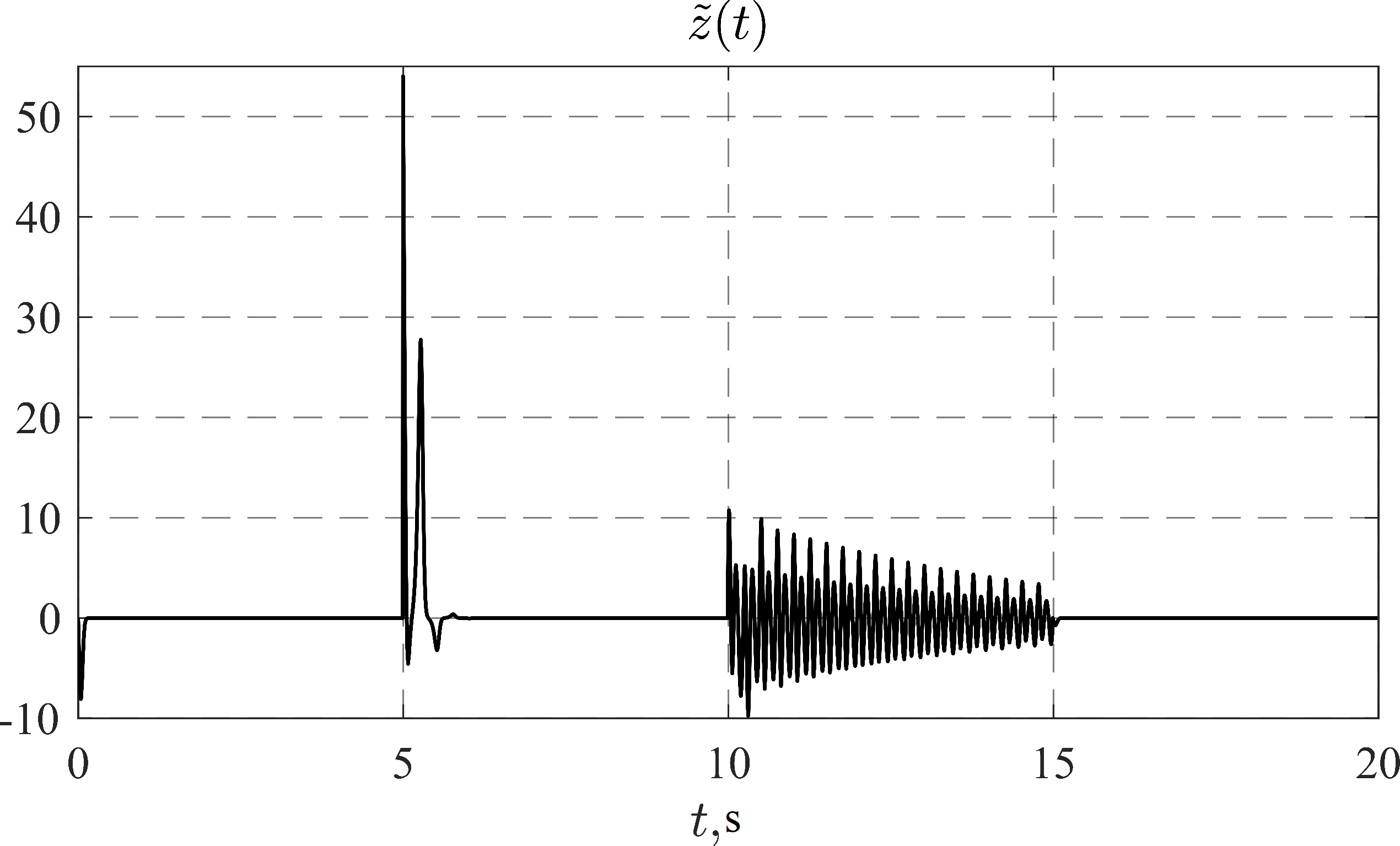} \\ c)}
\caption{Transient curves of $\tilde z\left( t \right)$ of the laws~\eqref{3.5} – (a),~\eqref{2.7} – (b) and~\eqref{2.3} – (c).}
\label{ris:image8}
\end{figure}

The transients that are depicted in Figure 8 validate that the tracking error $\tilde \theta (t)$ recovered exponenially to its equilibrium, as it was is proved in Theorem 3, when \linebreak $\overline \varphi \left( t \right) \in {\rm{s\text{-}PE}}$ and Assumption 3 was met.

Figure 9 presents the behaviour of the norm of $\tilde \Theta \left( t \right)$.

\begin{figure}[h!]
\center
\includegraphics[width=0.5\linewidth]{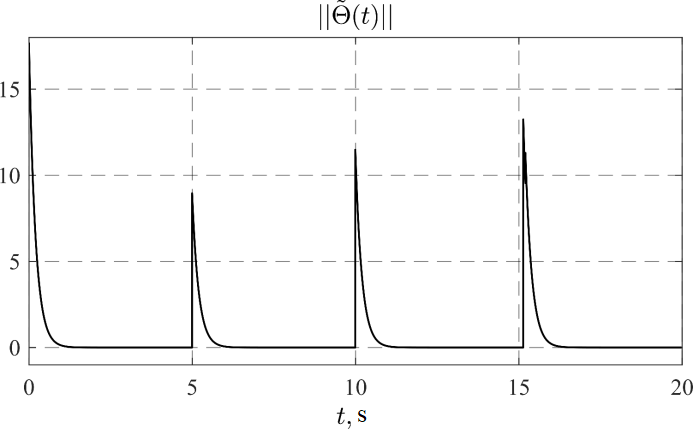}
\caption{Transient curve of the error $\tilde \Theta \left( t \right)$ norm.}
\label{ris:image9}
\end{figure}

Having analyzed Figure 9, it was concluded that the parameter error $\tilde \Theta \left( t \right)$ recovered to its equilibrium point when $\overline \varphi \left( t \right) \in {\rm{s\text{-}PE}}$ and Assumption 3 was met, which validated the conclusions made in Theorem 3.

As the number of the rank switches was finite $j \leqslant {j_{{\text{max}}}} < \infty $, then, according to Theorem 3 and Corollary 5, exponential recovery of $\tilde z\left( t \right)$ and $\tilde \Theta \left( t \right)$ to their respective equilibrium points was equivalent to exponential stability.

Figure 10 presents transients of $\left\| {\tilde \theta \left( t \right)} \right\|$ when the laws~\eqref{3.5},~\eqref{2.7} and~\eqref{2.3} were applied.

\begin{figure}[h!]
\center
\includegraphics[width=0.5\linewidth]{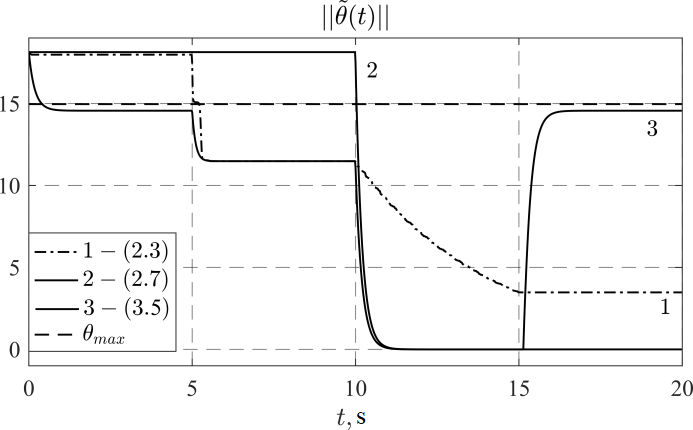}
\caption{Transient curves of $\left\| {\tilde \theta \left( t \right)} \right\|$ of the laws~\eqref{3.5},~\eqref{2.7} and~\eqref{2.3}.}
\label{ris:image10}
\end{figure}

The transients of $\left\| {\tilde \theta \left( t \right)} \right\|$ obtained with the help of the law~\eqref{3.5} confirmed the conclusions made in Theorem 1. The goal~\eqref{2.2} was achieved when $\overline \varphi \left( t \right) \in {\rm{s\text{-}FE}}$ and sufficient conditions were met, and $\tilde \theta \left( t \right)$ did exponentially converge to the set with the bound ${\theta _{{\text{max}}}}$, while such properties were ensured by~\eqref{2.3} only for all $t \geqslant 5$, and by~\eqref{2.7} -- only for $\overline \varphi \left( t \right) \in {\text{FE}}$.

Then it was set that ${\theta _0} = {{\begin{bmatrix}
  0&{ - 10}&{14} 
\end{bmatrix}}^{\text{\rm T}}}$, which did not meet the sufficient convergence conditions of Theorem 1 since $\left\| {\tilde \theta \left( {t_r^ + } \right)} \right\| \approx 4.9$ and \linebreak ${\theta _{{\text{max}}}} = \left\| \theta \right\| \approx 15$. Figure 11 shows the transients of $\left\| {\tilde \theta \left( t \right)} \right\|$ obtained under such choice of initial conditions when the laws~\eqref{3.5},~\eqref{2.7} and~\eqref{2.3} were applied.

\begin{figure}[h!]
\center
\includegraphics[width=0.5\linewidth]{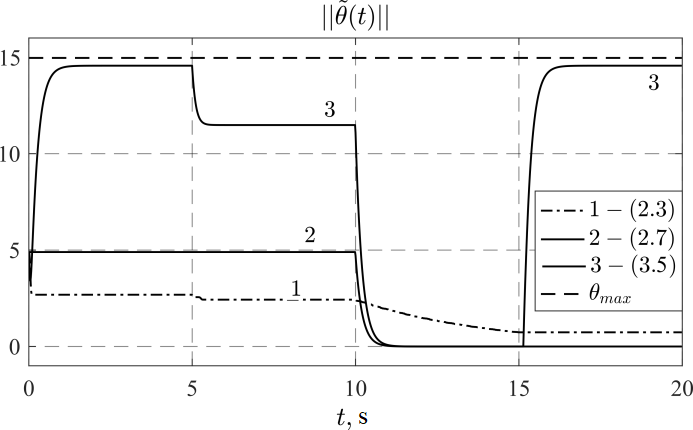}
\caption{Transient curves of $\left\| {\tilde \theta \left( t \right)} \right\|$ of the laws~\eqref{3.5},~\eqref{2.7} and~\eqref{2.3}.}
\label{ris:image11}
\end{figure}

The simulation results shown in Fig. 11 follows the results of Theorem 1. Indeed, when $\left\| {\tilde \theta \left( {t_r^ + } \right)} \right\| < {\theta _{{\text{max}}}}$, law~\eqref{3.5} did not converge (when $\left\| {\tilde \theta \left( {t_r^ + } \right)} \right\| = {\theta _{{\text{max}}}}$, it was quasi-convergent), and the error norm $\left\| {\tilde \theta \left( t \right)} \right\|$ could become greater than $\left\| {\tilde \theta \left( {t_r^ + } \right)} \right\|$.

\subsubsection{Second experiment}
The regression equation~\eqref{2.1} was defined as:
\begin{gather} \label{4.2.4}
\begin{gathered}
  z\left( t \right) = {{\overline \varphi }^{\text{\rm T}}}\left( t \right)\theta  = {\begin{bmatrix}
  {{{\overline \varphi }_1}\left( t \right)}&{{{\overline \varphi }_2}\left( t \right)}&{{{\overline \varphi }_3}\left( t \right)} 
\end{bmatrix}}{\begin{bmatrix}
  4 \\ 
  { - 8} \\ 
  {12} 
\end{bmatrix}};\\ 
  {{\overline \varphi }_1}\left( t \right) = \left\{ \begin{gathered}
   - 2{e^{ - t}}\cos \left( t \right){\text{, 0}} \leqslant t \leqslant 1 \hfill \\
  {e^{ - t}}{\text{, 1}} < t \leqslant 2 \hfill \\
  {e^{ - t}}\cos \left( t \right){\text{, }}t > 2 \hfill \\ 
\end{gathered}  \right.{\text{; }}{{\overline \varphi }_2}\left( t \right) = \left\{ \begin{gathered}
  {e^{ - t}}\cos \left( t \right){\text{, 0}} \leqslant t \leqslant 1 \hfill \\
   - 2{e^{ - t}}\cos \left( t \right){\text{, 1}} < t \leqslant 2 \hfill \\
  {e^{ - t}}{\text{ + 0}}{\text{,1}}{\text{, }}t > 2 \hfill \\ 
\end{gathered}  \right.{\text{;}} \\ 
  {\text{ }}{{\overline \varphi }_3}\left( t \right) = \left\{ \begin{gathered}
  {e^{ - t}}{\text{, 0}} \leqslant t \leqslant 1 \hfill \\
  {e^{ - t}}\cos \left( t \right){\text{, 1}} < t \leqslant 2 \hfill \\
   - 2{e^{ - t}}\cos \left( t \right){\text{, }}t > 2 \hfill \\ 
\end{gathered}  \right.. \\ 
\end{gathered}
\end{gather}

The parameters of the filter~\eqref{2.4}, algorithm of the eigenvalue virtual substitution~\eqref{3.2} and identification laws~\eqref{2.3},~\eqref{3.5} were set as:
\begin{gather} \label{4.2.5}
\begin{gathered}
l = 100,{\text{ }}\varepsilon  = 0,4,{\text{ }}\overline \varepsilon  = {10^{ - 10}}{\text{, }}{\gamma _0} = 5,\;{\gamma _1} = 1,{\text{ }}\Gamma  = {I_3}.
\end{gathered}
\end{gather}

In order to provide the same convergence rate for the laws~\eqref{3.5} and~\eqref{2.7}, the adaptive gain $\gamma$ of the law~\eqref{2.7} was defined similarly to~\eqref{3.5}, following the method of the regressor excitation normalization \cite{18}:
\begin{gather} \label{4.2.6}
\begin{gathered}
\gamma \left( t \right) = \left\{ \begin{gathered}
  {\gamma _1}{\text{, if  }}\omega \left( t \right) \leqslant {\text{min}}\left\{ {\lambda _{{\text{min}}}^n\left( t \right){\text{, }}{\varepsilon ^n}} \right\} \hfill \\
  \tfrac{{{\gamma _0}}}{{{\omega ^2}\left( t \right)}}{\text{ otherwise }} \hfill \\ 
\end{gathered}  \right..
\end{gathered}
\end{gather}

First of all, it was shown that the convergence conditions of the laws~\eqref{2.3},~\eqref{2.7} and~\eqref{3.5} were met. Figure 12 presents the behaviour of the disturbance $d\left( t \right)$ and regressor $\varphi \left( t \right)$ rank in the course of the experiment.

\begin{figure}[h!]
\begin{minipage}[h]{0.49\linewidth}
\center{\includegraphics[width=1\linewidth]{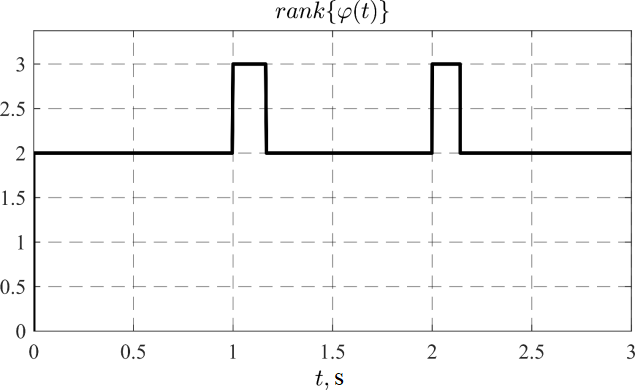} \\ a)}
\end{minipage}
\hfill
\begin{minipage}[h]{0.49\linewidth}
\center{\includegraphics[width=1\linewidth]{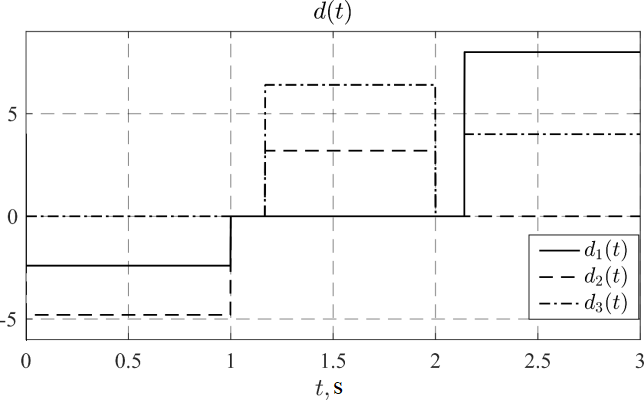} \\ b)}
\end{minipage}
\caption{Rank of the regressor $\varphi \left( t \right)$ (a), the disturbance $d\left( t \right)$ (b).}
\label{ris:image12}
\end{figure}

The time ranges $\left[ 1{\text{; 1}}{\text{,165}} \right]$ and $\left[ {{\text{2; 2}}{\text{,14}}} \right]$, at which $\rm{rank}\left\{{\varphi\left( t \right)} \right\} = 3$, were substantially shorter than the time intervals, when $\rm{rank}\left\{{\varphi \left( t \right)} \right\} = 2$. Therefore, unlike the experiment in Section 4.2.1, in this one the rank of the regressor was time-invariant almost everywhere. The rank differed from two when $\left[ {1{\text{; 1}}{\text{,165}}} \right]$ and $\left[ {{\text{2; 2}}{\text{,14}}} \right]$ as the filter~\eqref{2.4} mixed information about regressors with different bases. Considering~\eqref{2.7}, the convergence condition was satisfied over $\left[ {1{\text{; 1}}{\text{,165}}} \right]$ and $\left[ {{\text{2; 2}}{\text{,14}}} \right]$ due to the mixing effect.

In turn, for the law~\eqref{3.5} the necessary condition of convergence was satisfied for all $t \geqslant 0$. According to Fig. 12, the number of parameter switches was finite $j \leqslant {j_{{\text{max}}}} < \infty $ and $r \geqslant 1$, and then, by Assumption 3 and the results of Theorem 3 and Corollary 5, for~\eqref{3.5} all necessary and sufficient conditions of exponential convergence of the errors $\tilde z\left( t \right)$ and $\tilde \Theta \left( t \right)$ to zero were satisfied. Moreover, since
\begin{displaymath}
\forall t \in \left[ {0{\text{; 1}}} \right]{\text{\;}}{d_3}\left( t \right) = 0,{\text{\;}}\forall t \in \left[ {{\text{1; 2}}} \right]{\text{\;}}{d_1}\left( t \right) = 0,{\text{\;}}\forall t \in \left[ {2{\text{; 3}}} \right]{\text{\;}}{d_2}\left( t \right) = 0,
\end{displaymath}
then the conditions of partial identifiability described in Proposition 2 were also met in the course of the experiment.

Having verified that the convergence conditions were met, the experiments were conducted using the algorithms~\eqref{3.5},~\eqref{2.7} and~\eqref{2.3} under different initial conditions.

Firstly, it was set that ${\theta _0} = {{\begin{bmatrix}
  0&{ - 10}&{14} 
\end{bmatrix}}^{\text{\rm T}}}$, so the convergence conditions from Theorem 1 were not met since $\left\| {\tilde \theta \left( {t_r^ + } \right)} \right\| \approx 4,9$, while ${\theta _{\max }} = \left\| \theta  \right\| \approx 15$.

Figure 13 depicts the transients of the errors ${\tilde \theta _i}\left( t \right)$ for~\eqref{3.5} – (a),\linebreak \eqref{2.7} – (b) and~\eqref{2.3} – (c).

\begin{figure}[h!]
\begin{minipage}[h]{0.49\linewidth}
\center{\includegraphics[width=1\linewidth]{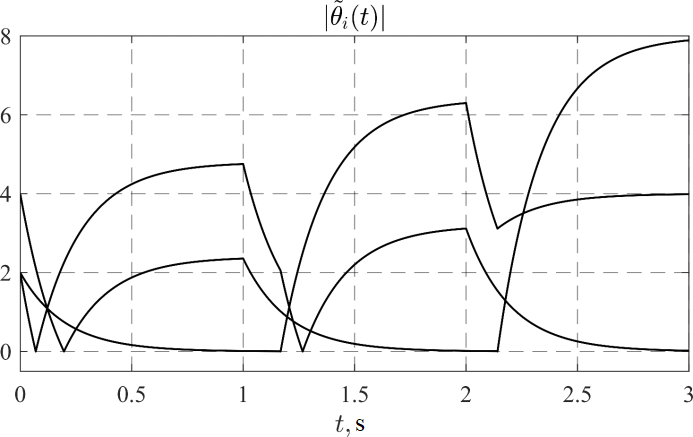} \\ а)}
\end{minipage}
\hfill
\begin{minipage}[h]{0.49\linewidth}
\center{\includegraphics[width=1\linewidth]{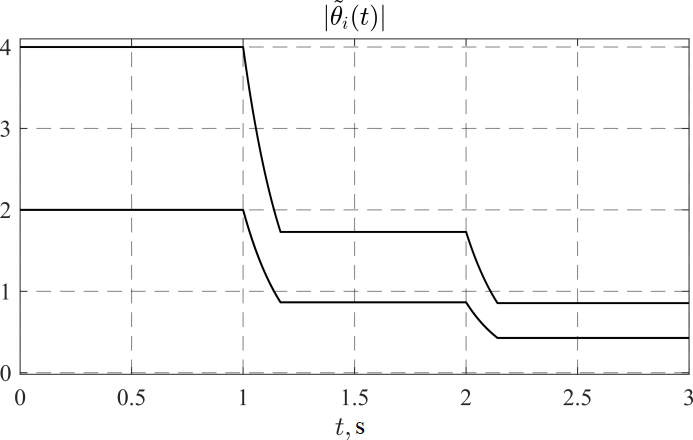} \\ b)}
\end{minipage}
\vfill
\center{\includegraphics[width=0.5\linewidth]{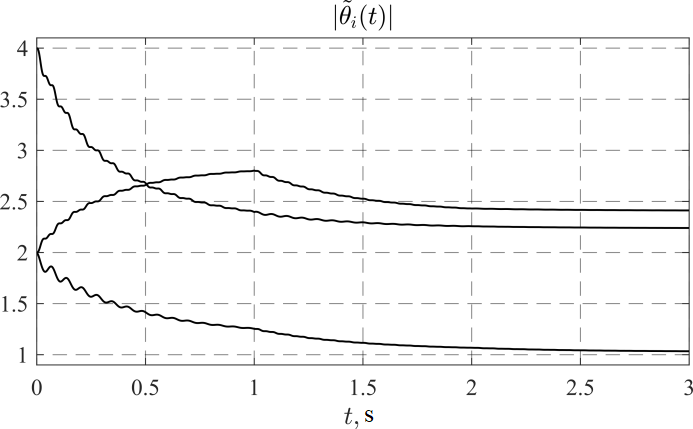} \\ c)}
\caption{Transient curves of the errors ${\tilde \theta _i}\left( t \right)$ of the laws~\eqref{3.5} – (a),~\eqref{2.7} – (b) and~\eqref{2.3} – (c).}
\label{ris:image13}
\end{figure}

The obtained transients confirmed the theoretical conclusions made in Remark 4. Indeed, under the conditions of the second statement of Proposition 2, the law~\eqref{3.5}, in contrast to~\eqref{2.7} and~\eqref{2.3}, provided a monotonic decrease of the error ${\tilde \theta _i}\left( t \right)$ over the corresponding time intervals when ${d_i}\left( t \right) = 0$:
\begin{displaymath}
    \left| {{{\tilde \theta }_3}\left( 1 \right)} \right| \leqslant \beta \left| {{{\tilde \theta }_3}\left( 0 \right)} \right|{\text{, }}\left| {{{\tilde \theta }_1}\left( 2 \right)} \right| \leqslant \beta \left| {{{\tilde \theta }_1}\left( 1 \right)} \right|{\text{, }}\left| {{{\tilde \theta }_2}\left( 3 \right)} \right| \leqslant \beta \left| {{{\tilde \theta }_2}\left( 2 \right)} \right|{\text{, }}\beta  \in \left( {0{\text{; 1}}} \right).
\end{displaymath}

Comparing the transients, the advantages of the law~\eqref{3.5} is seen over~\eqref{2.7} and~\eqref{2.3}. As for~\eqref{2.3}, the law~\eqref{3.5} ensured the first-order type transients of ${\tilde \theta _i}\left( t \right){\text{\;}}\forall i \in \overline {1,n}$ throughout the experiment. Compared to~\eqref{2.7}, the law~\eqref{3.5} converged not just over the time ranges $\left[ {1{\text{;1}}{\text{,165}}} \right]$ and $\left[ {\text{2; 2}{\text{,14}}} \right]$, but for all $t \geqslant 0$.

Figure 14 depicts the transients of $\tilde z\left( t \right)$ for laws~\eqref{3.5} - (a),~\eqref{2.7} - (b) and~\eqref{2.3} - (c).

\begin{figure}[h!]
\begin{minipage}[h]{0.49\linewidth}
\center{\includegraphics[width=1\linewidth]{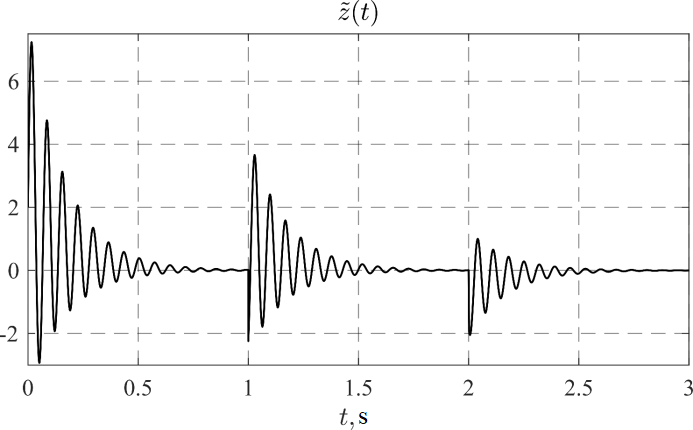} \\ a)}
\end{minipage}
\hfill
\begin{minipage}[h]{0.49\linewidth}
\center{\includegraphics[width=1\linewidth]{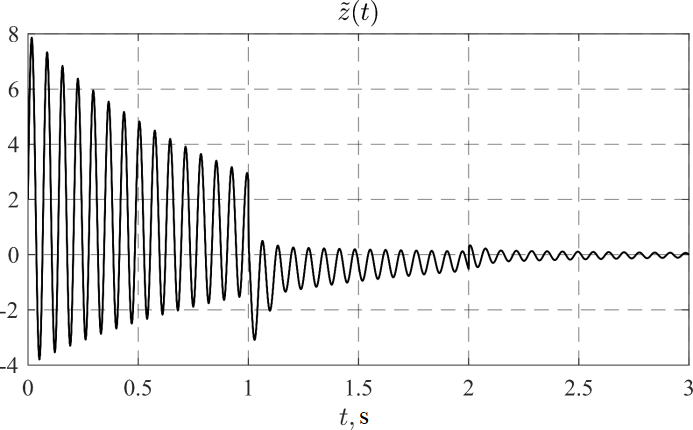} \\ b)}
\end{minipage}
\vfill
\center{\includegraphics[width=0.5\linewidth]{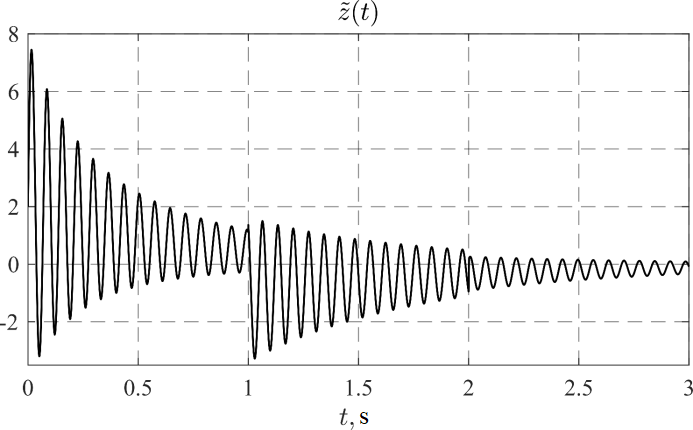} \\ c)}
\caption{Transient curves of the error $\tilde z\left( t \right)$ of the laws~\eqref{3.5} – (a),~\eqref{2.7} – (b) and~\eqref{2.3} – (c).}
\label{ris:image14}
\end{figure}

The transients in Fig. 14 confirm the exponential recovery of the tracking error $\tilde z\left( t \right)$ to its equilibrium point proved in Theorem 3 when $\overline \varphi \left( t \right) \in {\rm{s \text{-} PE}}$ and Assumption 3 was met.

Figure 15 shows transient curve of the $\tilde \Theta \left( t \right)$ norm.

\begin{figure}[h!]
\center
\includegraphics[width=0.5\linewidth]{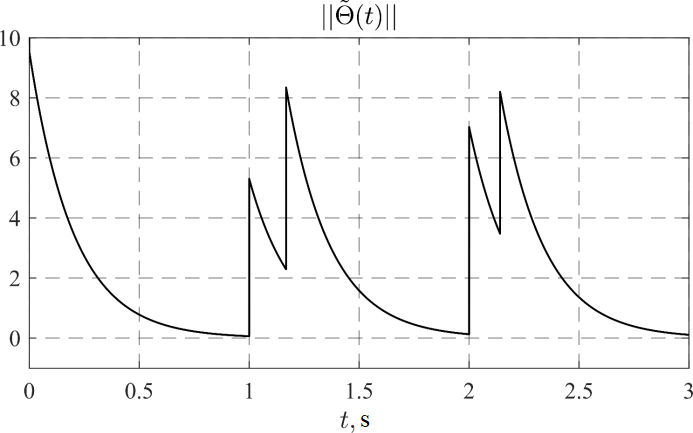}
\caption{Transient curve of the error $\tilde \Theta \left( t \right)$ norm.}
\label{ris:image15}
\end{figure}

Figure 15 validates the exponential recovery of the parameter error $\tilde \Theta \left( t \right)$ to its equilibrium point when $\overline \varphi \left( t \right) \in {\rm{s\text{-} PE}}$ and Assumption 3 was met, which followed the conclusions made in Theorem 3.

Since the number of rank switches was finite $j \leqslant {j_{{\text{max}}}} = 4 < \infty $, then according to the results of Corollary 5 the exponential recovery of the errors $\tilde z\left( t \right)$ and $\tilde \Theta \left( t \right)$ to their equilibrium points is equivalent to exponential stability.

Figure 16 presents transients of $\left\| {\tilde \theta \left( t \right)} \right\|$ for the laws~\eqref{3.5},~\eqref{2.7} and~\eqref{2.3}.

\begin{figure}[h!]
\center
\includegraphics[width=0.5\linewidth]{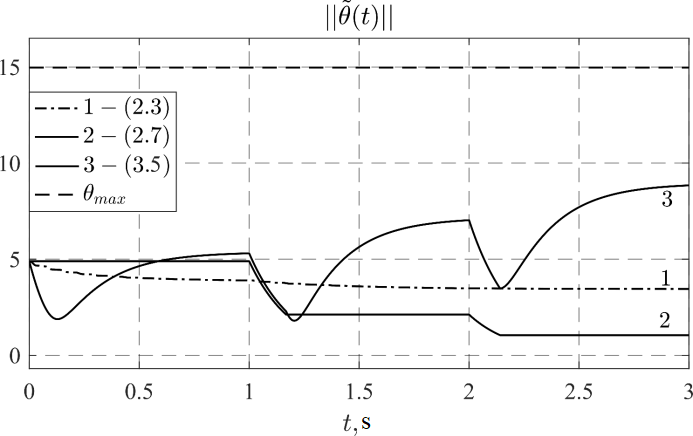}
\caption{Transient curves of $\left\| {\tilde \theta \left( t \right)} \right\|$ of the laws~\eqref{3.5},~\eqref{2.7} and~\eqref{2.3}.}
\label{ris:image16}
\end{figure}

The simulation results shown in Figure 16 validate the conclusions made in Theorem 1. Indeed, when $\left\| {\tilde \theta \left( {t_r^ + } \right)} \right\| < {\theta _{{\text{max}}}}$, the law~\eqref{3.5} was not convergent (when $\left\| {\tilde \theta \left( {t_r^ + } \right)} \right\| = {\theta _{{\text{max}}}}$, it was quasi-convergent), and the error norm $\left\| {\tilde \theta \left( t \right)} \right\|$ could become greater than $\left\| {\tilde \theta \left( {t_r^ + } \right)} \right\|$.

Then it was set that ${\theta _0} = {{\begin{bmatrix}
  0&5&0 
\end{bmatrix}}^{\text{\rm T}}}$, which, according to Theorem 1, ensured that sufficient conditions of convergence of the law~\eqref{3.5} were met:
\begin{displaymath}
{\beta _1} = \left\| {\tilde \theta \left( {t_r^ + } \right)} \right\|{\left\| \theta \right\|^{ - 1}} \approx \tfrac{{18}}{{15}} = 1,2{\text{; }}\tfrac{1}{{{\beta _1}}} + {e^{ - {\gamma _0}\delta }} = \tfrac{1}{{1,2}} + {e^{ - 5 \cdot 1}} \approx 0.84 \in \left( {0{\text{; 1}}} \right).    
\end{displaymath}
	
Figure 17 shows the transients of $\left\| {\tilde \theta \left( t \right)}\right\|$ obtained under such initial conditions, when the laws~\eqref{3.5},~\eqref{2.7} and~\eqref{2.3} were applied.

\begin{figure}[h!]
\center
\includegraphics[width=0.5\linewidth]{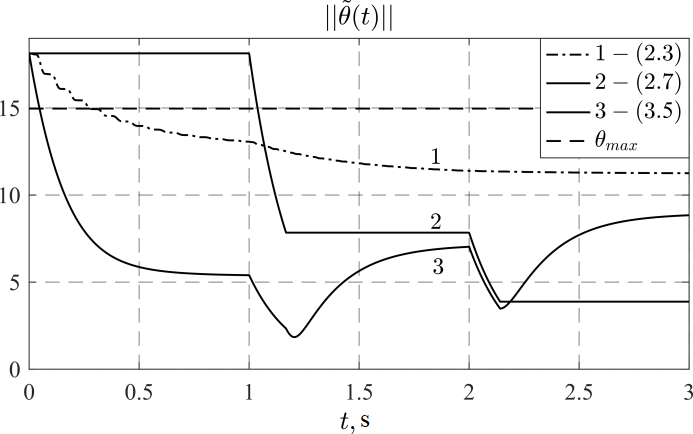}
\caption{Transient curves of $\left\| {\tilde \theta \left( t \right)} \right\|$ of the laws~\eqref{3.5},~\eqref{2.7} and~\eqref{2.3}.}
\label{ris:image17}
\end{figure}

The transient of $\left\| {\tilde \theta \left( t \right)} \right\|$ for the law~\eqref{3.5} confirmed the conclusions made in Theorem 1. The goal~\eqref{2.2} was achieved when $\overline \varphi \left( t \right) \in {\rm{s\text{-}FE}}$ and sufficient conditions were met, and $\tilde \theta \left( t \right)$ did exponentially converge to the set with the bound ${\theta _{{\text{max}}}}$. Considering~\eqref{2.7}, such properties held only when $\overline \varphi \left( t \right) \in {\text{FE}}$.

Thus, the numerical experiments confirmed all theoretically stated properties of the proposed law~\eqref{3.5}. The results of Section 3.1 are valid in the general case $\overline \varphi \left( t \right) \in {\rm{s\text{-} FE}}$, and the results of Sections 3.2 and 3.3 are applicable under Assumptions 2 and 3, respectively.

\newpage

\section{Conclusion}
In order to solve the identification problem of the unknown time-invariant parameters of a linear regression equation under the regressor semi-finite excitation, a procedure of dynamic regressor extension, regularization and mixing was proposed that generalized the well-known DREM method and extended the area of its applicability as far as practical scenarios were concerned.

In contrast to the conventional gradient-based identification law~\eqref{2.3}, the proposed procedure provided element-wise monotonicity of errors when Assumption 2 was met and exponential convergence of the tracking error of the function~\eqref{2.1} when the regressor was semi-persistently exciting with the rank not less than one.

In contrast to the conventional DREM procedure, the developed one, firstly, relaxed the requirement of the regressor finite excitation previously required for convergence of~\eqref{2.7} and ensured that the unknown parameters identification error decreased when the weaker condition of the regressor semi-finite excitation was met, and secondly, guaranteed exponential convergence of the regressand~\eqref{2.1} tracking error when the regressor was semi-persistently exciting with the rank not less than one.

The scope of future research is the analysis and development of the dynamic regressor extension, regularization, and mixing procedure to solve the following problems:
\begin{enumerate}
\item[--] synthesis of the adaptive control schemes with relaxed requirements of the regressor excitation to ensure exponential convergence of the reference model tracking error;
\item[--] development of adaptive state observers with relaxed regressor excitation requirements for exponential convergence of plant states tracking error to zero;
\item[--] using partial identifiability conditions (see Proposition 2 and Fig. 13,a, Fig. 7,a, Fig. 2,a) to identify the full vector of plant unknown parameters in case of over-parameterization;
\item[--] based on Proposition 2 and simulation results (see Fig. 13,a, Fig. 7,a, Fig. 2,a), development of an identification law that does not require a finite or persistent excitation of the regressor to provide exponential convergence of the identification error of the full vector of unknown parameters.
\end{enumerate}
\newpage
\appendix{}
 \renewcommand{\theequation}{\mbox{A.}\arabic{equation}}
\begin{proofofproposition}{\ref{st1}} The lower bounds of the regressor  $\omega \left( t \right)$ are written on the basis of Corollaries 1-4:
\begin{gather*}
\begin{gathered}
  \overline \varphi \left( t \right) \in {\text{PE}} \Leftrightarrow \forall t \geqslant kT{\text{ }}\omega \left( t \right) = {\text{det}}\left\{ {\Phi \left( t \right)} \right\} = \prod\limits_{i = 1}^n {{\lambda _i}\left( t \right)}  \geqslant \lambda _{{\text{min}}}^n\left( t \right) > {\mu ^n} > 0, \hfill \\
  \overline \varphi \left( t \right) \in {\text{FE}} \Leftrightarrow \forall t \in \left[ {{t_\delta }{\text{; }}{t_\delta } + \delta } \right] \subset \left[ {t_r^ + {\text{; }}{t_e}} \right]{\text{ }}\omega \left( t \right) = \prod\limits_{i = 1}^n {{\lambda _i}\left( t \right)}  \geqslant \lambda _{{\text{min}}}^n\left( t \right) > {\mu ^n} > 0, \hfill \\
  \overline \varphi \left( t \right) \in {\rm{s\text{-} PE}} \Leftrightarrow \forall t \geqslant kT{\text{ }}\omega \left( t \right) = {\varepsilon ^{\overline r}}\prod\limits_{i = 1}^r {{\lambda _i}\left( t \right)}  \geqslant {\text{min}}\left\{ {\lambda _{{\text{min}}}^n\left( t \right){\text{, }}{\varepsilon ^n}} \right\} > 0, \hfill \\
  \overline \varphi \left( t \right) \in {\rm{s\text{-} FE}} \Leftrightarrow \forall t \in \left[ {{t_\delta }{\text{; }}{t_\delta } + \delta } \right] \subset \left[ {t_r^ + {\text{; }}{t_e}} \right]{\text{ }}\omega \left( t \right) = {\varepsilon ^{\overline r}}\prod\limits_{i = 1}^r {{\lambda _i}\left( t \right)}  \geqslant {\text{min}}\left\{ {\lambda _{{\text{min}}}^n\left( t \right){\text{, }}{\varepsilon ^n}} \right\} > 0. \hfill \\ 
\end{gathered}
\end{gather*}
as was to be proved in Proposition 1.
\end{proofofproposition}

\begin{proofoftheorem}{\ref{th:1}}
\textbf{1.} As, following Corollaries 1 and 2, the following implications hold when ${{\overline \varphi \left( t \right) \in {\text{FE}}} \mathord{\left/
 {\vphantom {{\overline \varphi \left( t \right) \in {\text{FE}}} {\overline \varphi \left( t \right) \in {\text{PE}}}}} \right.
 \kern-\nulldelimiterspace} {\overline \varphi \left( t \right) \in {\text{PE}}}}$:
 \begin{gather}\label{A.1}
 \begin{gathered}
  \overline \varphi \left( t \right) \in {\text{PE}} \Leftrightarrow \forall t \geqslant kT{\text{ }}{\lambda _{{\text{min}}}}\left( t \right) > \mu  > 0, \\ 
  \overline \varphi \left( t \right) \in {\text{FE}} \Leftrightarrow \forall t \in \left[ {{t_\delta }{\text{; }}{t_\delta } + \delta } \right] \subset \left[ {t_r^ + {\text{; }}{t_e}} \right]{\text{ }}{\lambda _{{\text{min}}}}\left( t \right) > \mu  > 0,
\end{gathered}
 \end{gather}
then, when ${{\overline \varphi \left( t \right) \in {\text{FE}}} \mathord{\left/
 {\vphantom {{\overline \varphi \left( t \right) \in {\text{FE}}} {\overline \varphi \left( t \right) \in {\text{PE}}}}} \right.
 \kern-\nulldelimiterspace} {\overline \varphi \left( t \right) \in {\text{PE}}}}$, in accordance with \eqref{3.2}, zero eigenvalues in $\Lambda \left( t \right)$ are not substituted $\Xi \left( t \right) = {0_{n \times n}}$, the equality $\Phi \left( t \right) = \varphi \left( t \right)$ holds for the regressor matrix $\Phi \left( t \right)$, then it holds for the unknown parameters $\Theta $ that $\Theta  = \theta $ owing to ${\overline \Lambda ^{ - 1}}\left( t \right)\Xi \left( t \right) = {0_{n \times n}}$, and the identification law \eqref{3.5} coincides with \eqref{2.7} up to the definition of the adaptive gain $\gamma$, from which it follows that \eqref{3.5} ensures $b_1$--$b_5$ when ${{\overline \varphi \left( t \right) \in {\text{FE}}} \mathord{\left/
 {\vphantom {{\overline \varphi \left( t \right) \in {\text{FE}}} {\overline \varphi \left( t \right) \in {\text{PE}}}}} \right.
 \kern-\nulldelimiterspace} {\overline \varphi \left( t \right) \in {\text{PE}}}}$.

\textbf{2.} The following function, in which time arguments are omitted for the sake of brevity, is introduced:
\begin{gather}\label{A.2}
 \begin{gathered}
\forall t \in \left[ {t_r^ + {\text{; }}{t_e}} \right]{\text{\;}}L = {\tilde \theta ^{\text{T}}}\tilde \theta.
\end{gathered}
 \end{gather}

The equation~\eqref{A.2} is differentiated along the solutions of~\eqref{3.5} to obtain:
\begin{gather}\label{A.3}
 \begin{gathered}
\dot L =  - 2{\tilde \theta ^{\text{T}}}\left( {\gamma \omega \left( {\omega \hat \theta  - \omega \theta  + \omega V{{\overline \Lambda }^{ - 1}}\Xi {V^{\text{T}}}\theta } \right)} \right) =  - 2{\tilde \theta ^{\text{T}}}\gamma {\omega ^2}\tilde \theta  - 2{\tilde \theta ^{\text{T}}}\gamma {\omega ^2}V{\overline \Lambda ^{ - 1}}\Xi {V^{\text{T}}}\theta.
\end{gathered}
 \end{gather}

Considering Assumption 1 and the definition of $\gamma$, the upper bound of~\eqref{A.3} for all $t \in \left[ {{t_\delta }{\text{; }}{t_\delta } + \delta } \right] \subset \left[ {t_r^ + {\text{; }}{t_e}} \right]$ is written as:
\begin{gather}\label{A.4}
 \begin{gathered}
\dot L \leqslant  - 2{{\tilde \theta }^{\text{T}}}\tfrac{{{\gamma _0}}}{{{\omega ^2}}}{\omega ^2}\tilde \theta  - 2{{\tilde \theta }^{\text{T}}}\tfrac{{{\gamma _0}}}{{{\omega ^2}}}{\omega ^2}V{{\overline \Lambda }^{ - 1}}\Xi {V^{\text{T}}}\theta  \leqslant  \\ 
   \leqslant  - 2{{\tilde \theta }^{\text{T}}}{\gamma _0}\tilde \theta  - 2{{\tilde \theta }^{\text{T}}}{\gamma _0}V{{\overline \Lambda }^{ - 1}}\Xi {V^{\text{T}}}\theta  \leqslant  - 2{\gamma _0}{\left\| {\tilde \theta } \right\|^2} + 2{\gamma _0}\left\| {\tilde \theta } \right\|{\theta _{{\text{max}}}}.
\end{gathered}
 \end{gather}

Here spectral norm of the multiplier $V{\overline \Lambda ^{ - 1}}\Xi {V^{\text{T}}}$, which value is one as the matrices $V$ and $V^{\rm T}$ are orthogonal ones, is calculated to obtain~\eqref{A.4}.

Assuming that $a = \sqrt {2{\gamma _0}} \left\| {\tilde \theta } \right\|{\text{,\;}}$$b = \sqrt {2{\gamma _0}} {\theta _{{\text{max}}}}{\text{}}$ and using the inequality \linebreak $ - {a^2} + ab \leqslant  - \tfrac{1}{2}{a^2} + \tfrac{1}{2}{b^2}$, it is obtained from~\eqref{A.4}:
\begin{gather}\label{A.5}
 \begin{gathered}
\dot L \leqslant  - {\gamma _0}{\left\| {\tilde \theta } \right\|^2} + {\gamma _0}\theta _{{\text{max}}}^2.
\end{gathered}
 \end{gather}
 
The solution of the differential inequality~\eqref{A.5} for all $t \in \left[ {{t_\delta }{\text{; }}{t_\delta } + \delta } \right]$ is written as:
\begin{gather}\label{A.6}
 \begin{gathered}
\forall t \in \left[ {{t_\delta }{\text{; }}{t_\delta } + \delta } \right]{\text{ }}L \leqslant {e^{ - {\gamma _0}\left( {t - {t_\delta }} \right)}}{\left\| {\tilde \theta \left( {{t_\delta }} \right)} \right\|^2} + \theta _{{\text{max}}}^2.
\end{gathered}
 \end{gather}

Considering~\eqref{A.6}, $L = {\left\| {\tilde \theta } \right\|^2}$ and the fact that for all $c,{\text{\;}}d$ the inequalities \linebreak $\sqrt {{c^2} + {d^2}}  \leqslant \sqrt {{c^2}} + \sqrt {{d^2}} $ hold, we obtain:
\begin{gather}\label{A.7}
 \begin{gathered}
\left\| {\tilde \theta \left( {{t_\delta } + \delta } \right)} \right\| \leqslant {e^{ - 0,5{\gamma _0}\delta }}\left\| {\tilde \theta \left( {{t_\delta }} \right)} \right\| + {\theta _{{\text{max}}}}.
\end{gathered}
 \end{gather}
 
As for the most conservative case, it holds that $\omega \left( t \right) \equiv 0$ for all \linebreak $t \in \left\{ {\left[ {t_r^ + {\text{; }}{t_\delta }} \right]{\text{, }}\left[ {{t_\delta } + \delta {\text{; }}{t_e}} \right]} \right\}$, therefore, the inequalities $\left\| {\tilde \theta \left( {t_r^ + } \right)} \right\| \geqslant \left\| {\tilde \theta \left( {{t_\delta }} \right)} \right\|{\text{ }}$, \linebreak $\left\| {\tilde \theta \left( {{t_e}} \right)} \right\| \leqslant \left\| {\tilde \theta \left( {{t_\delta } + \delta } \right)} \right\|$ also hold, using which~\eqref{A.7} is rewritten as:
\begin{gather}\label{A.8}
 \begin{gathered}
\left\| {\tilde \theta \left( {{t_e}} \right)} \right\| \leqslant {e^{ - 0,5{\gamma _0}\delta }}\left\| {\tilde \theta \left( {t_r^ + } \right)} \right\| + {\theta _{{\text{max}}}}.
\end{gathered}
 \end{gather}

The premise 2.1) is substituted into~\eqref{A.8} to obtain:
\begin{gather}\label{A.9}
 \begin{gathered}
\left\| {\tilde \theta \left( {{t_e}} \right)} \right\| \leqslant \left( {{e^{ - 0,5{\gamma _0}\delta }} + \tfrac{1}{{{\beta _1}}}} \right)\left\| {\tilde \theta \left( {t_r^ + } \right)} \right\|.
\end{gathered}
 \end{gather}
 
Hence, the choice of ${\gamma _0}$ on the basis of the condition
\begin{gather}\label{A.10}
 \begin{gathered}
0 < {e^{ - 0,5{\gamma _0}\delta }} + \tfrac{1}{{{\beta _1}}} < 1 \Leftrightarrow {\gamma _0} > \tfrac{{ - 2{\text{ln}}\left( {1 - \tfrac{1}{{{\beta _1}}}} \right)}}{\delta }
\end{gathered}
 \end{gather}
allows one to ensure that the premise 2.2) also holds and, as a consequence, obtain the following:
\begin{gather}\label{A.11}
 \begin{gathered}
\left\| {\tilde \theta \left( {{t_e}} \right)} \right\| \leqslant \underbrace {\left( {{e^{ - 0,5{\gamma _0}\delta }} + \tfrac{1}{{{\beta _1}}}} \right)}_{0 < \beta  < 1}\left\| {\tilde \theta \left( {t_r^ + } \right)} \right\|{\text{,}}
\end{gathered}
 \end{gather}
which means that the error $\tilde \theta \left( t \right)$ decreases over the time range $\left[ {t_r^ + {\text{;\;}}{t_e}} \right]$.

The substitution of~\eqref{A.11} into the upper bound of $\tilde z\left( {{t_e}} \right)$ yields:
\begin{gather}\label{A.12}
 \begin{gathered}
\left| {\tilde z\left( {{t_e}} \right)} \right| \leqslant {\overline \varphi _{{\text{max}}}}\left\| {\tilde \theta \left( {{t_e}} \right)} \right\| \leqslant {\overline \varphi _{{\text{max}}}}\beta \left\| {\tilde \theta \left( {t_r^ + } \right)} \right\| = \beta \left| {\tilde z\left( {t_r^ + } \right)} \right|{\text{,}}
\end{gathered}
 \end{gather}
which completes the proof of the second statement and verifies the convergence of~\eqref{3.5} when $\overline \varphi \left( t \right) \in {\rm{s \text{-} FE}}$ and the premises 2.1) and 2.2) hold.

\textbf{3.} The derivative of $\tilde \Theta \left( t \right)$ is calculated to prove the third statement:
\begin{gather}\label{A.13}
 \begin{gathered}
\dot {\tilde {\Theta}} \left( t \right) =  - \gamma \left( t \right){\omega ^2}\left( t \right)\tilde \Theta \left( t \right) - \dot \Theta \left( t \right){\text{.}}
\end{gathered}
 \end{gather}

The general solution of the differential equation~\eqref{A.13} is:
\begin{gather}\label{A.14}
 \begin{gathered}
\tilde \Theta \left( t \right) = \phi \left( {t,{\text{ }}{t_0}} \right)\tilde \Theta \left( {{t_0}} \right) - \int\limits_{{t_0}}^t {\phi \left( {t,{\text{ }}\tau } \right)\dot \Theta \left( \tau  \right)d\tau } {\text{,}}
\end{gathered}
 \end{gather}
where $\phi \left( {t,s} \right) = {e^{ - \int\limits_s^t {\gamma \left( \tau  \right){\omega ^2}\left( \tau  \right)d\tau } }}.$

As, owing to $\sqrt {{\gamma _1}}  \notin {L_2}{\text{, }}\tfrac{{\sqrt {{\gamma _0}} }}{{\omega \left( t \right)}} \notin {L_2}$ and $\omega \left( t \right) \notin {L_2}$, for all possible switches of the nonlinear operator in~\eqref{3.5} it is true that $\sqrt \gamma \omega \left( t \right) \notin {L_2}$, then the function $\phi \left( {t,s} \right)$ has the following properties:
\begin{gather}\label{A.15}
 \begin{gathered}
\sqrt \gamma  \omega \left( t \right) \notin {L_2} \Leftrightarrow \left\{ \begin{gathered}
  0 < \phi \left( {t,s} \right) \leqslant 1, \hfill \\
  {\text{li}}{{\text{m}}_{t \to \infty }}\phi \left( {t,s} \right) = 0. \hfill \\ 
\end{gathered}  \right.
\end{gathered}
 \end{gather}

Using the first property, the upper bound of~\eqref{A.14} is obtained:
\begin{gather}\label{A.16}
 \begin{gathered}
\tilde \Theta \left( t \right) \leqslant \phi \left( {t,{\text{ }}{t_0}} \right)\tilde \Theta \left( {{t_0}} \right) - \Theta \left( t \right).
\end{gathered}
 \end{gather}

On the basis of~\eqref{A.16} and definitions $\tilde \Theta \left( t \right) = \tilde \theta \left( t \right){\text{ + }}d\left( t \right){\text{,\;}}\Theta \left( t \right) = \theta  - d\left( t \right)$ we have:
\begin{gather}\label{A.17}
 \begin{gathered}
\tilde \theta \left( t \right) \leqslant \phi \left( {t,{\text{ }}{t_0}} \right)\tilde \Theta \left( {{t_0}} \right) - \theta.
\end{gathered}
 \end{gather}

From this, based on the second property of~\eqref{A.15}, it follows that \linebreak ${\text{li}}{{\text{m}}_{t \to \infty }}\left\| {\tilde \theta \left( t \right)} \right\| \leqslant {\theta _{{\text{max}}}}{\text{,}}$ which completes the proof of the third statement of the theorem.

\textbf{4.} When the condition $\overline \varphi \left( t \right) \in {\rm{s\text{-}PE}}$ is met, in accordance with the third statement of Proposition 1 for all $t \geqslant kT$ it holds that $\omega \left( t \right) \geqslant {\text{min}}\left\{ {\lambda _{{\text{min}}}^n\left( t \right){\text{, }}{\varepsilon ^n}} \right\} > 0$ and, consequently, the function $\phi \left( {t{\text{, }}kT} \right)$ is written as:
\begin{gather}\label{A.18}
 \begin{gathered}
\phi \left( {t,{\text{ }}kT} \right) = {e^{ - {\gamma _0}\left( {t - kT} \right)}}.
\end{gathered}
 \end{gather}

Then, having solved~\eqref{A.13} for all $t \geqslant kT$, the following is obtained in a similar manner to \eqref{A.14}-\eqref{A.17}:
\begin{gather}\label{A.19}
 \begin{gathered}
\left\| {\tilde \theta \left( t \right)} \right\| \leqslant {e^{ - {\gamma _0}\left( {t - kT} \right)}}\left\| {\tilde \Theta \left( {kT} \right)} \right\| + {\theta _{{\text{max}}}}.
\end{gathered}
 \end{gather}
from which it follows that, when $\overline \varphi \left( t \right) \in {\rm{s\text{-}PE}}$, the errors $\tilde \theta \left( t \right)$ exponentially convergence to the set with the bound ${\theta _{{\text{max}}}}$, which completes the proof of the theorem.

\end{proofoftheorem}

\begin{proofoftheorem}{\ref{th:2}} \textbf{I.} To prove the first statement of Theorem 2, the equation~\eqref{3.4} is written in the element-wise form:
\begin{gather}\label{A.20}
 \begin{gathered}
{\Upsilon _i}\left( t \right) = \omega \left( t \right){\Theta _i}{\text{, }}\forall i \in \left\{ {1, \ldots ,n} \right\}.
\end{gathered}
 \end{gather}

Given~\eqref{A.20}, the law~\eqref{3.5} for all $i \in \left\{{1, \ldots ,n} \right\}$ is written as follows:
\begin{gather}\label{A.21}
 \begin{gathered}
{\dot {\hat \theta} _i}\left( t \right) = {\dot {\tilde \Theta} _i}\left( t \right) =  - \gamma \left( t \right)\omega \left( t \right)\left( {\omega \left( t \right){{\hat \theta }_i}\left( t \right) - \omega \left( t \right){\Theta _i}} \right){\text{ = }} - \gamma \left( t \right){\omega ^2}\left( t \right){\tilde \Theta _i}\left( t \right){\text{.}}
\end{gathered}
\end{gather}

As $\gamma \left( t \right){\omega ^2}\left( t \right) > 0$, then ${\text{sign}}\left\{ {{{\dot {\tilde \Theta} }_i}\left( t \right)} \right\} = {\text{const}}$, and it holds for ${\tilde \Theta _i}\left( t \right)$ that \linebreak $\left| {{{\tilde \Theta }_i}\left( {{t_a}} \right)} \right| \leqslant \left| {{{\tilde \Theta }_i}\left( {{t_b}} \right)} \right|{\text{ }}\forall {t_a} \geqslant {t_b}$, which was to be proved in part I of the theorem.

\textbf{II.} When $\overline \varphi \left( t \right) \in {\rm{s\text{-}FE}}$ and Assumption 2 is met, in accordance with Corollary 4 the solution of the equation~\eqref{A.13} over $\left[ {{t_\delta }{\text{; }}{t_\delta } + \delta } \right]$ is written as:
\begin{gather}\label{A.22}
 \begin{gathered}
\tilde \Theta \left( t \right) = \phi \left( {t,{\text{ }}{t_\delta }} \right)\tilde \Theta \left( {{t_\delta }} \right) = {e^{ - {\gamma _0}\left( {t - {t_\delta }} \right)}}\tilde \Theta \left( {{t_\delta }} \right).
\end{gathered}
 \end{gather}

Considering the most conservative case, for all $t \in \left\{ {\left[ {t_r^ + {\text{; }}{t_\delta }} \right]{\text{, }}\left[ {{t_\delta } + \delta {\text{; }}{t_e}} \right]} \right\}$ it holds that $\omega \left( t \right) \equiv 0$, therefore we have the inequalities $\left\| {\tilde \Theta \left( {t_r^ + } \right)} \right\| \geqslant \left\| {\tilde \Theta \left( {{t_\delta }} \right)} \right\|{\text{, }}\left\| {\tilde \Theta \left( {{t_e}} \right)} \right\| \leqslant \left\| {\tilde \Theta \left( {{t_\delta } + \delta } \right)} \right\|$, on the base of which the upper bound of $\tilde \Theta \left( t \right)$ at the time instant ${t_e}$ is obtained:
\begin{gather}\label{A.23}
 \begin{gathered}
\left\| {\tilde \Theta \left( {{t_e}} \right)} \right\| \leqslant {e^{ - {\gamma _0}\delta }}\left\| {\tilde \Theta \left( {t_r^ + } \right)} \right\|.
\end{gathered}
 \end{gather}
 
The definition $\beta  = {e^{ - {\gamma _0}\delta }} \in \left( {0{\text{; 1}}} \right)$ is introduced into~\eqref{A.23} to complete the proof that the error $\tilde \Theta \left( t \right)$ decreases over $\left[ {t_r^ + {\text{; }}{t_e}} \right]$.

To prove the error $\tilde z\left( t \right)$ reduction, the correctness of the following implication owing to $V_1^{\text{T}}\left( t \right){V_2} = {0_{r \times \overline r}}$ is taken into consideration:
\begin{gather}\label{A.24}
 \begin{gathered}
y\left( t \right) = \varphi \left( t \right)\theta  = {V_1}\left( t \right){\Lambda _1}\left( t \right)V_1^{\text{T}}\left( t \right)\left( {\theta  - {V_2}V_2^{\text{T}}\theta } \right) = \\ = \varphi \left( t \right)\left( {\theta  - {V_2}V_2^{\text{T}}\theta } \right) = \varphi \left( t \right)\Theta  =  \\ 
   = \int\limits_{t_0^ + }^t {{e^{ - l\left( {t - \tau } \right)}}\overline \varphi \left( \tau  \right){{\overline \varphi }^{\text{T}}}\left( \tau  \right)d\tau } \Theta  = \int\limits_{t_0^ + }^t {{e^{ - l\left( {t - \tau } \right)}}\overline \varphi \left( \tau  \right)z\left( \tau  \right)d\tau }  =  \\ 
   = \int\limits_{t_0^ + }^t {{e^{ - l\left( {t - \tau } \right)}}\overline \varphi \left( \tau  \right)\underbrace {{{\overline \varphi }^{\text{T}}}\left( \tau  \right)\theta }_{z\left( \tau  \right)}d\tau }  = \int\limits_{t_0^ + }^t {{e^{ - l\left( {t - \tau } \right)}}\overline \varphi \left( \tau  \right)\underbrace {{{\overline \varphi }^{\text{T}}}\left( \tau  \right)\Theta }_{z\left( \tau  \right)}d\tau }  \\ 
   \Updownarrow  \\ 
  z\left( t \right) = {{\overline \varphi }^{\text{T}}}\left( t \right)\theta  = {{\overline \varphi }^{\text{T}}}\left( t \right)\left( {\theta  - {V_2}V_2^{\text{T}}\theta } \right) = {{\overline \varphi }^{\text{T}}}\left( t \right)\Theta .
\end{gathered}
 \end{gather}
 
Then, considering~\eqref{A.22}, the upper bound of the tracking error is written as:
\begin{gather}\label{A.25}
 \begin{gathered}
\forall t \in \left[ {{t_\delta }{\text{; }}{t_\delta } + \delta } \right]{\text{ }}\left| {\tilde z\left( t \right)} \right| \leqslant {\overline \varphi _{{\text{max}}}}{e^{ - {\gamma _0}\left( {t - {t_\delta }} \right)}}\left\| {\tilde \Theta \left( {{t_\delta }} \right)} \right\|,
\end{gathered}
 \end{gather}
from which, owing to~\eqref{A.23}, we immediately have:
\begin{gather}\label{A.26}
 \begin{gathered}
\left| {\tilde z\left( {{t_e}} \right)} \right| \leqslant {\overline \varphi _{{\text{max}}}}\beta \left\| {\tilde \Theta \left( {t_r^ + } \right)} \right\| = \beta \left| {\tilde z\left( {t_r^ + } \right)} \right|{\text{,}}
\end{gathered}
 \end{gather}
which was to be proved in part II.

\textbf{III.} When Assumption 2 is met, for all $t \in \left[ {{t_0}{\text{;\;}}\infty } \right)$ the solution of the error~\eqref{A.13} is written as:
\begin{gather}\label{A.27}
 \begin{gathered}
\tilde \Theta \left( t \right) = \phi \left( {t,{\text{ }}{t_0}} \right)\tilde \Theta \left( {{t_0}} \right){\text{,}}
\end{gathered}
 \end{gather}
from which, according to the second property of~\eqref{A.15}, it follows that:
\begin{gather}\label{A.28}
 \begin{gathered}
\sqrt {\gamma \left( t \right)} \omega \left( t \right) \notin {L_2} \Leftrightarrow {\text{li}}{{\text{m}}_{t \to \infty }}\left\| {\tilde \Theta \left( t \right)} \right\| = 0,
\end{gathered}
 \end{gather}
which holds for all possible variants of switches of the nonlinear operator~\eqref{3.5} owing to $\sqrt {{\gamma _1}}  \notin {L_2}{\text{, }}\tfrac{{\sqrt {{\gamma _0}} }}{{\omega \left( t \right)}} \notin {L_2}$ 
and $\omega \left( t \right) \notin {L_2}$.

Having applied the implication~\eqref{A.28} to obtain the upper bound of~\eqref{A.24}, we have:
\begin{gather}\label{A.29}
 \begin{gathered}
\sqrt {\gamma \left( t \right)} \omega \left( t \right) \notin {L_2} \Leftrightarrow {\text{li}}{{\text{m}}_{t \to \infty }}\left| {\tilde z\left( t \right)} \right| \leqslant {\text{li}}{{\text{m}}_{t \to \infty }}\left( {{{\overline \varphi }_{{\text{max}}}}\left\| {\tilde \Theta \left( t \right)} \right\|} \right) = 0.
\end{gathered}
 \end{gather}

Thus, all statements of the third part of Theorem 2 are proved.

\textbf{IV.} When $\overline \varphi \left( t \right) \in {\rm{s\text{-} PE}}$, then \eqref{A.18} holds $\forall t \geqslant kT$, and therefore the following bound is obtained on the basis of~\eqref{A.22}:
\begin{gather}\label{A.30}
 \begin{gathered}
\forall t \geqslant kT{\text{ }}\left\| {\tilde \Theta \left( t \right)} \right\| \leqslant {e^{ - {\gamma _0}\left( {t - kT} \right)}}\left\| {\tilde \Theta \left( {kT} \right)} \right\|{\text{,}}
\end{gathered}
 \end{gather}
which proves the exponential convergence of the error $\tilde \Theta \left( t \right)$ to zero for all $t \geqslant kT$.

Having~\eqref{A.30} at hand, considering the boundedness of $\left\| {\overline \varphi \left( t \right)} \right\| \leqslant {\overline \varphi _{{\text{max}}}}$ and using~\eqref{A.24}, the exponential convergence of the error $\tilde z\left( t \right)$ for all $t \geqslant kT$ can be proved in the similar way to~\eqref{A.25}, which completes the proof of Theorem 2.

\end{proofoftheorem}

\begin{proofoftheorem}{\ref{th:3}} When $\overline \varphi \left( t \right) \in {\rm{s\text{-} PE}}$, on the basis of the third statement of proved Proposition 1 for all $t \geqslant kT$ $\omega \left( t \right) \geqslant {\text{min}}\left\{ {\lambda _{{\text{min}}}^n\left( t \right){\text{, }}{\varepsilon ^n}} \right\} > 0$ holds, and therefore the equation~\eqref{A.13} is written as:
\begin{gather}\label{A.31}
 \begin{gathered}
\forall t \geqslant kT{\text{\;}}\dot {\tilde {\Theta}} \left( t \right) =  - {\gamma _0}\tilde \Theta \left( t \right) - \dot \Theta \left( t \right).
\end{gathered}
 \end{gather}

Owing to Assumption 3, the derivative $\dot \Theta \left( t \right)$ is written as follows according to~\eqref{3.7}:
\begin{gather}\label{A.32}
 \begin{gathered}
\dot \Theta \left( t \right) = \sum\limits_{j = 1}^\infty  {{\Delta _j}\delta \left( {t - {t_j}} \right)}.
\end{gathered}
 \end{gather}

Considering~\eqref{A.32}, the solution of the differential equation~\eqref{A.31} is obtained:
\begin{gather}\label{A.33}
 \begin{gathered}
\forall t \geqslant kT{\text{ }}\tilde \Theta \left( t \right) = {e^{ - {\gamma _0}\left( {t - kT} \right)}}\tilde \Theta \left( {kT} \right) - \int\limits_{kT}^t {{e^{ - {\gamma _0}\left( {t - \tau } \right)}}\sum\limits_{j = 1}^\infty  {{\Delta _j}\delta \left( {\tau  - {t_j}} \right)} } d\tau.
\end{gathered}
 \end{gather}
 
Following the sifting property of the Dirac function, for any differentiable function $f\left( t \right)$ we have:
\begin{gather}\label{A.34}
\begin{gathered}
  \int\limits_{{t_0}}^t {f\left( \tau  \right)\delta \left( {\tau  - {t_j}} \right)} {\text{ }}d\tau  = \left. {f\left( {{t_j}} \right)h\left( {\tau  - {t_j}} \right)} \right|_{{t_0}}^t =  \\ 
   = f\left( {{t_j}} \right)h\left( {t - {t_j}} \right) - f\left( {{t_j}} \right)\underbrace {h\left( {{t_0} - {t_j}} \right)}_{ = 0} \equiv f\left( {{t_j}} \right)h\left( {t - {t_j}} \right). \\ 
\end{gathered}
 \end{gather}

On the basis of~\eqref{A.34} the equation~\eqref{A.33} is rewritten as:
\begin{gather}\label{A.35}
\begin{gathered}
\forall t \geqslant kT{\text{ }}\tilde \Theta \left( t \right) = {e^{ - {\gamma _0}\left( {t - kT} \right)}}\tilde \Theta \left( {kT} \right) - \sum\limits_{j = 1}^\infty  {{e^{ - {\gamma _0}\left( {t - {t_j}} \right)}}{\Delta _j}h\left( {t - {t_j}} \right)}.
\end{gathered}
 \end{gather}
 
Having multiplied~\eqref{A.35} by ${\tilde \Theta ^{\text{T}}}\left( {kT} \right)$, it is obtained:
\begin{gather}\label{A.36}
\begin{gathered}
\forall t \geqslant kT{\text{\;}}{\tilde \Theta ^{\text{T}}}\left( {kT} \right)\tilde \Theta \left( t \right) = {e^{ - {\gamma _0}\left( {t - kT} \right)}}{\left\| {\tilde \Theta \left( {kT} \right)} \right\|^2} - \\-\sum\limits_{j = 1}^\infty  {{e^{ - {\gamma _0}\left( {t - {t_j}} \right)}}{{\tilde \Theta }^{\text{T}}}\left( {kT} \right){\Delta _j}h\left( {t - {t_j}} \right)}.
\end{gathered}
 \end{gather}

The term ${e^{ - {\gamma _0}\left( {t - kT} \right)}}{\left\| {\tilde \Theta \left( {kT} \right)} \right\|^2}$ is put outside the brackets in the right-hand side of the equation~\eqref{A.36} to obtain for all $t \geqslant kT$ that:
\begin{gather}\label{A.37}
\begin{array}{c}
{{\tilde \Theta }^{\rm{T}}}\left( {kT} \right)\tilde \Theta \left( t \right) = \underbrace {\left( {1 - {\textstyle{1 \over {{{\left\| {\tilde \Theta \left( {kT} \right)} \right\|}^2}}}}\sum\limits_{j = 1}^\infty  {{e^{ - {\gamma _0}\left( {kT - {t_j}} \right)}}{{\tilde \Theta }^{\rm{T}}}\left( {kT} \right){\Delta _j}h\left( {t - {t_j}} \right)} } \right)}_{ \in R} \times \\
 \times {e^{ - {\gamma _0}\left( {t - kT} \right)}}{\rm{ }}{{\tilde \Theta }^{\rm{T}}}\left( {kT} \right)\tilde \Theta \left( {kT} \right),\\
\tilde \Theta \left( t \right) = \left( {1 - {\textstyle{1 \over {{{\left\| {\tilde \Theta \left( {kT} \right)} \right\|}^2}}}}\sum\limits_{j = 1}^\infty  {{e^{ - {\gamma _0}\left( {kT - {t_j}} \right)}}{{\tilde \Theta }^{\rm{T}}}\left( {kT} \right){\Delta _j}h\left( {t - {t_j}} \right)} } \right){e^{ - {\gamma _0}\left( {t - kT} \right)}}{\rm{ }}\tilde \Theta \left( {kT} \right),
\end{array}
 \end{gather}
where $\left\| {\tilde \Theta \left( {kT} \right)} \right\| \ne 0$ since for all $t \in \left[ {{t_0}{\text{; }}kT} \right){\text{ }}\omega \left( t \right) \equiv 0 \Rightarrow \dot {\hat {\theta}} \left( t \right) = 0 \Rightarrow \left\| {\tilde \Theta \left( {kT} \right)} \right\| \geqslant \\ \geqslant \left\| {\tilde \Theta \left( {{t_0}} \right)} \right\|.$

The equation~\eqref{A.37} allows one to have the first expression from~\eqref{3.8} up to the following notation:
\begin{gather}\label{A.38}
\begin{gathered}
a\left( {{t_j}} \right) = \left| {1 - \tfrac{1}{{{{\left\| {\tilde \Theta \left( {kT} \right)} \right\|}^2}}}\sum\limits_{j = 1}^\infty  {{e^{ - {\gamma _0}\left( {kT - {t_j}} \right)}}{{\tilde \Theta }^{\text{T}}}\left( {kT} \right){\Delta _j}h\left( {t - {t_j}} \right)} } \right|.
\end{gathered}
 \end{gather}

So the exponential recovery of the parameter error $\tilde \Theta \left( t \right)$ to its equilibrium point is proved.

Having~\eqref{A.24} at hand, the upper bound of the tracking error $\left| {\tilde z\left( t \right)} \right|$ is written as:
\begin{gather}\label{A.39}
\begin{gathered}
\forall t \geqslant kT{\text{\;}}\left| {\tilde z\left( t \right)} \right| \leqslant a\left( {{t_j}} \right){\overline \varphi _{{\text{max}}}}{e^{ - {\gamma _0}\left( {t - kT} \right)}}\left\| {\tilde \Theta \left( {kT} \right)} \right\| = a\left( {{t_j}} \right){e^{ - {\gamma _0}\left( {t - kT} \right)}}\left| {\tilde z\left( {kT} \right)} \right|.
\end{gathered}
 \end{gather}

Therefore, the exponential recovery of the error $\tilde z\left( t \right)$ to its equilibrium point is also proved.

If, additionally, for $a\left( {{t_j}} \right)$ there exists an upper bound $a_{{{\text{max}}}}$, then it is immediately obtained from~\eqref{3.8} that:
\begin{gather}\label{A.40}
{\rm{ }}\left\{ {\begin{array}{*{20}{c}}
{\mathop {{\rm{lim}}}\limits_{t \to \infty } \left\| {\tilde \Theta \left( t \right)} \right\| \le \mathop {{\rm{lim}}}\limits_{t \to \infty } \left( {{a_{{\rm{max}}}}{e^{ - {\gamma _0}\left( {t - kT} \right)}}\left\| {\tilde \Theta \left( {kT} \right)} \right\|} \right) = 0}\\
\begin{array}{l}
\mathop {{\rm{lim}}}\limits_{t \to \infty } \left| {\tilde z\left( t \right)} \right| \le \mathop {{\rm{lim}}}\limits_{t \to \infty } \left( {{a_{{\rm{max}}}}{{\overline \varphi }_{{\rm{max}}}}{e^{ - {\gamma _0}\left( {t - kT} \right)}}\left\| {\tilde \Theta \left( {kT} \right)} \right\|} \right) = \\
 = \mathop {{\rm{lim}}}\limits_{t \to \infty } \left( {{a_{{\rm{max}}}}{e^{ - {\gamma _0}\left( {t - kT} \right)}}\left| {\tilde z\left( {kT} \right)} \right|} \right) = 0
\end{array}
\end{array}} \right..
 \end{gather}

Hence, the tracking error $\tilde z\left( t \right)$ and the parameter error $\tilde \Theta \left( t \right)$ are exponentially stable, which completes the proof of Theorem 3.

\end{proofoftheorem}

\begin{proofofcorollary}{\ref{cor5}} According to the first statement of Corollary 5, it is assumed that the number of $\Theta \left( t \right)$ changes is finite: $j \leqslant {j_{{\text{max}}}} < \infty $.

Then the following upper bound of the function $a\left( {{t_j}} \right)$ is obtained:
\begin{gather}\label{A.41}
\begin{array}{c}
a\left( {{t_j}} \right) = \left| {1 - {\textstyle{1 \over {{{\left\| {\tilde \Theta \left( {kT} \right)} \right\|}^2}}}}\sum\limits_{j = 1}^{{j_{{\rm{max}}}}} {{e^{ - {\gamma _0}\left( {kT - {t_j}} \right)}}{{\tilde \Theta }^{\rm{T}}}\left( {kT} \right){\Delta _j}h\left( {t - {t_j}} \right)} } \right| \le \\
 \le 1 + \left| {{\textstyle{1 \over {{{\left\| {\tilde \Theta \left( {kT} \right)} \right\|}^2}}}}\sum\limits_{j = 1}^{{j_{{\rm{max}}}}} {{e^{ - {\gamma _0}\left( {kT - {t_j}} \right)}}{{\tilde \Theta }^{\rm{T}}}\left( {kT} \right){\Delta _j}h\left( {t - {t_j}} \right)} } \right| \le \\
 \le 1 + {\textstyle{1 \over {\left\| {\tilde \Theta \left( {kT} \right)} \right\|}}}\sum\limits_{j = 1}^{{j_{{\rm{max}}}}} {\left\| {{\Delta _j}} \right\|{e^{ - {\gamma _0}\left( {kT - {t_j}} \right)}}h\left( {t - {t_j}} \right)}.
\end{array}
 \end{gather}

As, when $j$ is finite, the number of time instants ${t_j}$ is also finite, then the exponential multiplier in the sum~\eqref{A.41} is bounded, and the following definition holds:
\begin{gather}\label{A.42}
\begin{gathered}
a\left( {{t_j}} \right) \leqslant 1 + \tfrac{1}{{\left\| {\tilde \Theta \left( {kT} \right)} \right\|}}\sum\limits_{j = 1}^{{j_{{\text{max}}}}} {\left\| {{\Delta _j}} \right\|{e^{ - {\gamma _0}\left( {kT - {t_j}} \right)}}h\left( {t - {t_j}} \right)}  = {a_{{\text{max}}}}, 
\end{gathered}
 \end{gather}
which was to be proved in the first part of the corollary.

To prove the second statement of the Corollary, the upper bound of $\left\| {{\Delta _j}} \right\|$ is taken into consideration, and the upper bound of $a\left( {{t_j}} \right)$ is obtained similarly to~\eqref{A.42}, but under the condition of the infinite number of switches:
\begin{gather}\label{A.43}
\begin{array}{c}
a\left( {{t_j}} \right) \le 1 + \left| {{\textstyle{1 \over {{{\left\| {\tilde \Theta \left( {kT} \right)} \right\|}^2}}}}\sum\limits_{j = 1}^\infty  {{e^{ - {\gamma _0}\left( {kT - {t_j}} \right)}}{{\tilde \Theta }^{\rm{T}}}\left( {kT} \right){\Delta _j}h\left( {t - {t_j}} \right)} } \right| \le \\
 \le 1 + \sum\limits_{j = 1}^\infty  {c\left( {{t_j}} \right)h\left( {t - {t_j}} \right)} .
\end{array}
 \end{gather}

The series from \eqref{A.43} is of positive terms, and all its subsums are bounded because of monotonicity $0 < c\left( {{t_{j + 1}}} \right) \leqslant c\left( {{t_j}} \right)$, and therefore $1 + \sum\limits_{j = 1}^\infty  {c\left( {{t_j}} \right)h\left( {t - {t_j}} \right)}  \leqslant {a_{{\text{max}}}}$, which completes the proof of Corollary 5.

\end{proofofcorollary}

\begin{proofofproposition}{\ref{st2}} As, when ${{\overline \varphi \left( t \right) \in {\text{FE}}} \mathord{\left/
 {\vphantom {{\overline \varphi \left( t \right) \in {\text{FE}}} {\overline \varphi \left( t \right) \in {\text{PE}}}}} \right.
 \kern-\nulldelimiterspace} {\overline \varphi \left( t \right) \in {\text{PE}}}}$, the following implications hold according to Corollaries 1 and 2:
 \begin{displaymath}
 \begin{gathered}
  \overline \varphi \left( t \right) \in {\text{PE}} \Leftrightarrow \forall t \geqslant kT{\text{ }}{\lambda _{{\text{min}}}}\left( t \right) > \mu  > 0, \\ 
  \overline \varphi \left( t \right) \in {\text{FE}} \Leftrightarrow \forall t \in \left[ {{t_\delta }{\text{; }}{t_\delta } + \delta } \right] \subset \left[ {t_r^ + {\text{; }}{t_e}} \right]{\text{ }}{\lambda _{{\text{min}}}}\left( t \right) > \mu  > 0, \\ 
\end{gathered}
 \end{displaymath}
then, when $\overline \varepsilon  = 0$, according to~\eqref{3.3} we have $\Xi \left( t \right) = {0_{n \times n}}$, as a result ${\overline \Lambda ^{ - 1}}\left( t \right)\Xi \left( t \right) = {0_{n \times n}}$ and, consequently, ${{\overline \varphi \left( t \right) \in {\text{FE}}} \mathord{\left/
 {\vphantom {{\overline \varphi \left( t \right) \in {\text{FE}}} {\overline \varphi \left( t \right) \in {\text{PE}}}}} \right.
 \kern-\nulldelimiterspace} {\overline \varphi \left( t \right) \in {\text{PE}}}} \Rightarrow d\left( t \right) = {0_n} \Rightarrow \Theta \left( t \right) = \theta $, which completes the proof of statement (а) of Proposition 2.
 
The necessity of conditions ${{\overline \varphi \left( t \right) \in {\rm{s{-}FE}}} \mathord{\left/
 {\vphantom {{\overline \varphi \left( t \right) \in {\rm{s\text{-} FE}}} {\overline \varphi \left( t \right) \in {\rm{s\text{-} PE}}}}} \right.
 \kern-\nulldelimiterspace} {\overline \varphi \left( t \right) \in {\rm{s\text{-} PE}}}}$ follows from the fact that only if $0 < r < n$, the premises of the statement b) are consistent ($\exists p > 0{\text{\;}}\sum\limits_{i = 1}^{n - p} {{w_i}{\varphi _i}\left( t \right)}  = {0_n}{\text{, }}{w_i} \ne 0$). The necessity of the condition $n > 2$ follows from the contradiction, which occurs when $n = 2$ in general case $\left( {{\varphi _1}\left( t \right) \ne {0_n}} \right)$:
\begin{displaymath}
{w_1}{\varphi _1}\left( t \right) + {w_2}{\varphi _2}\left( t \right) = {0_n}{\text{ }}{w_1} \ne 0,{\text{ }}{w_2} = 0.
\end{displaymath}  

The next step is to prove the necessity and sufficiency of the following condition to ensure that $\exists M \subset \left\{ {1,...,n} \right\}{\text{, }}\left| M \right| = p{\text{, }}\forall i \in M{\text{, }}{\Theta _i} = {\theta _i}$ :
\begin{gather}\label{A.44}
\begin{gathered}
\sum\limits_{i = 1}^{n - p} {{w_i}{\varphi _i}\left( t \right)}  + \sum\limits_{j = n - p + 1}^n {{w_j}{\varphi _j}\left( t \right)}  = {0_n}{\text{, }}{w_i} \ne 0,{\text{ }}{w_j} = 0.
\end{gathered}
 \end{gather}

\textbf{Necessity.} To begin with, it should be noted that according to~\eqref{3.5}, the elements of the vector of new unknown parameters $\Theta$ coincide with the elements of the vector of original parameters $\theta$ if the corresponding elements of the vector $d$ are equal to zero. Therefore, $d$ is considered in more detail. If $\overline r > 0$, the multiplication ${\overline \Lambda ^{ - 1}}\left( t \right)\Xi \left( t \right)$ has the following structure:
\begin{gather}\label{A.45}
\begin{gathered}
{\overline \Lambda ^{ - 1}}\left( t \right)\Xi \left( t \right) = {\begin{bmatrix}
  {\Lambda _1^{ - 1}\left( t \right)}&{{0_{r \times \overline r}}} \\ 
  {{0_{\overline r \times r}}}&{{\varepsilon ^{ - 1}}{I_{\overline r}}} 
\end{bmatrix}} {\begin{bmatrix}
  {{0_r}}&{{0_{r \times \overline r}}} \\ 
  {{0_{\overline r \times r}}}&{\varepsilon {I_{\overline r}}} 
\end{bmatrix}} = {\begin{bmatrix}
  {{0_r}}&{{0_{r \times \overline r}}} \\ 
  {{0_{\overline r \times r}}}&{{I_{\overline r}}} 
\end{bmatrix}}.
\end{gathered}
\end{gather}

Then, owing to the notation~\eqref{3.4}, the definition of $d$ is rewritten as:
\begin{gather}\label{A.46}
\begin{gathered}
d = V\left( t \right){\overline \Lambda ^{ - 1}}\left( t \right)\Xi \left( t \right){V^{\text{T}}}\left( t \right)\theta  = {V_2}V_2^{\text{T}}\theta  = {\left[ {{d_1} \ldots {d_i} \ldots {d_n}} \right]^{\text{T}}},
\end{gathered}
\end{gather}
from which it follows that $d$ has $p$ zero elements if, in particular, the number of zero rows and columns of the matrix ${V_2}V_2^{\text{T}}$ is $p$, which, in turn, is satisfied when the matrix ${V_2}$ has $p$ zero rows.

Following the definition of the singular decomposition of a positively semi-definite symmetric matrix~\cite{15,16}, the matrix ${V_2}$ can be obtained as a solution of a homogeneous system of linear algebraic equations:
\begin{gather}\label{A.47}
\begin{gathered}
\varphi \left( t \right)V_2^k = \sum\limits_{i = 1}^n {v_i^k{\varphi _i}\left( t \right)}  = {0_n}{\text{, }}\forall k \in \left\{ {1,{\text{ }}\overline r} \right\}{\text{,}}
\end{gathered}
\end{gather}
where $V_2^k$ is the $k^{\rm th}$ column of the matrix ${V_2}$.

To prove the necessity of the condition~\eqref{A.44}, it is to be shown that if ${w_j} \ne 0$, then the vector $V_2^k{\text{, }}\forall k \in \left\{ {1,{\text{ }}\overline r} \right\}{\text{,}}$ does not contain zero elements.

The expression~\eqref{A.47} can be rewritten in the following equivalent form (taking into account the orthonormality of $V_2^k{\text{, }}\forall k \in \left\{ {1,{\text{ }}\overline r} \right\}$):
\begin{gather}\label{A.48}
\begin{gathered}
  \varphi \left( t \right)V_2^k = \sum\limits_{i = 1}^n {v_i^k{\varphi _i}\left( t \right)}  = {\text{ }}\tfrac{1}{{\sqrt {\sum\limits_{i = 1}^n {w_i^2} } }}\sum\limits_{i = 1}^n {{w_i}{\varphi _i}\left( t \right)}  = \\ = \tfrac{1}{{\sqrt {\sum\limits_{i = 1}^n {w_i^2} } }}\left( {\sum\limits_{i = 1}^{n - p} {{w_i}{\varphi _i}\left( t \right)}  + \sum\limits_{j = n - p + 1}^n {{w_j}{\varphi _j}\left( t \right)} } \right) =  \hfill \\
   = \sum\limits_{i = 1}^{n - p} {v_i^k{\varphi _i}\left( t \right)}  + \sum\limits_{j = n - p + 1}^n {v_j^k{\varphi _j}\left( t \right)}  = {0_n}. \hfill \\ 
\end{gathered}
\end{gather}

Since we consider only nontrivial solutions to find $V_2^k$, if the condition~\eqref{A.44} is not satisfied, the set of solutions is given as follows:
\begin{displaymath}
\begin{gathered}
v_i^k = \tfrac{{{w_i}}}{{\sqrt {\sum\limits_{i = 1}^n {w_i^2} } }} \ne 0 {\rm ;\;} v_j^k = \tfrac{{{w_j}}}{{\sqrt {\sum\limits_{i = 1}^n {w_i^2} } }} \ne 0, 
\end{gathered}
\end{displaymath}
and then $V_2^k{\text{, }}\forall k \in \left\{ {1,{\text{ }}\overline r} \right\}{\text{,}}$ does not include zero elements and, consequently, \linebreak $\bcancel{\exists }{d_i} = 0 \Rightarrow \bcancel{\exists }M \subset \left\{ {1,...,n} \right\}{\text{, }}\left| M \right| = p{\text{, }}\forall i \in M{\text{, }}{\Theta _i} = {\theta _i}$, which completes the proof of necessity of the condition~\eqref{A.45}.

\textbf{Sufficiency.} Following the statement of the proposition, when the condition~\eqref{A.44} is met, the solution set of the equation of the form~\eqref{A.47} is defined as follows:
\begin{displaymath}
\begin{gathered}
v_i^k = \tfrac{{{w_i}}}{{\sqrt {\sum\limits_{i = 1}^n {w_i^2} } }} \ne 0;\;v_j^k = \tfrac{{{w_j}}}{{\sqrt {\sum\limits_{i = 1}^n {w_i^2} } }} = 0,
\end{gathered}
\end{displaymath}
and then the vector $V_2^k{\text{, }}\forall k \in \left\{ {1,{\text{ }}\overline r} \right\}{\text{,}}$ includes $p$ zero elements and, consequently, \linebreak $\exists M \subset \left\{ {1,...,n} \right\}{\text{, }}\left| M \right| = p{\text{, }}\forall i \in M{\text{, }}{\Theta _i} = {\theta _i}$, which completes the proof of sufficiency of the condition~\eqref{A.44}.

Thus, the condition~\eqref{A.44} is necessary and sufficient for the identifiability of $p$ elements of the unknown parameters vector $\theta$, which completes the proof of the second statement of Proposition 2.
 
\end{proofofproposition}

\end{document}